\documentclass{aastex}  

\usepackage{emulateapj5} 
\usepackage{graphics}

\begin{document}

\slugcomment{submitted to The Astronomical Journal}

\shorttitle{Broad-band photometry for galaxy evolution studies}
\shortauthors{Gil de Paz \& Madore}

\title{On the Optimization of Broad-Band Photometry for Galaxy Evolution Studies}

\author{A. Gil de Paz\altaffilmark{1,2} and B. F. Madore\altaffilmark{1,3}}

\altaffiltext{1} {NASA/IPAC Extragalactic Database, California Institute of Technology, MS 100-22, Pasadena, CA 91125; agpaz, barry@ipac.caltech.edu}
\altaffiltext{2} {Jet Propulsion Laboratory, MS 183-900, Pasadena, CA 91109}
\altaffiltext{3} {The Observatories, Carnegie Institution of Washington, 813 Santa Barbara Street, Pasadena, CA 91101}

\begin{abstract}

We have derived the uncertainties to be expected in the derivation of
galaxy physical properties (star formation history, age, metallicity,
reddening) when comparing broad-band photometry to the predictions of
evolutionary synthesis models. We have obtained synthetic colors for a
large sample ($\sim$9000) of artificial galaxies assuming different
star formation histories, ages, metallicities, reddening values, and
redshifts. The colors derived have been perturbed by adopting
different observing errors, and compared back to the evolutionary
synthesis models grouped in different sets. The comparison has been
performed using a combination of Monte Carlo simulations, a Maximum
Likelihood Estimator and Principal Component Analysis. After comparing
the input and derived output values we have been able to compute the
uncertainties and covariant degeneracies between the galaxy physical
properties as function of (1) the set of observables available, (2)
the observing errors, and (3) the galaxy properties themselves. In
this work we have considered different sets of observables, some of
them including the standard Johnson/Cousins ($UBVR_{C}I_{C}$) and {\it
Sloan Digital Sky Survey (SDSS)} bands in the optical, the {\it 2
Micron All Sky Survey (2MASS)} bands in the near-infrared, and the
{\it Galaxy Evolution Explorer (GALEX)} bands in the UV, at three
different redshifts, $z$=0.0, 0.7, and 1.4. This study is intended to
represent a basic tool for the design of future projects on galaxy
evolution, allowing an estimate of the optimal band-pass combinations
and signal-to-noise ratios required for a given scientific objective.
\end{abstract}


\section {Introduction}
\label{sec1}

The catalogs produced by wide-field and all-sky surveys currently
under development (e.g., GALEX; Martin et al$.$ 1997, SDSS; York et
al$.$ 2000, 2MASS; Skrutskie et al$.$ 1997, DENIS; Epchtein et al$.$
1997) in combination with astronomical databases like the NASA/IPAC
Extragalactic Database (NED) are beginning to provide easy access to
extensive cross-correlated UV, optical, and near-infrared (NIR)
photometry for millions of galaxies. The comparison of this huge
amount of photometric data with the predictions of state-of-the-art
galaxy population synthesis models provides an opportunity to obtain a
more complete picture of the evolution of galaxies. One obvious goal
is to gain insight into the star formation history, as well as the
chemical and dust content evolution of galaxies from the high and
intermediate redshift Universe down to the present.

However, the reliability and precision of the derived galaxy
properties expected to be found from these studies will depend on many
factors; among these are: (1) the number of bands and wavelength
coverage of the available observations, (2) the measuring
uncertainties, and (3) the degeneracies between the different galaxy
properties given the available photometric bands. In order to
discriminate between different scenarios of galaxy evolution, the
comparison of photometric data and evolutionary synthesis models
should also include the quantification of the uncertainties and
covariances between the galaxy properties derived. In addition, the
use of different external inputs (evolutionary tracks, stellar
atmospheres libraries, etc$.$) in the evolutionary synthesis models
leads to various discrepancies in the output results which may also
result in significant uncertainties in the properties derived. While,
a detailed study of these latter effects is beyond the scope of this
paper, a nice approach to this problem may be found in Charlot,
Worthey \& Bressan (1996; see also Bruzual 2000, Cervi\~no, Luridiana
\& Castander 2000, Cervi\~no et al$.$ 2001).

Some recent studies have incorporated the effects of the observing
errors and flux calibration uncertainties in the determination of the
properties of different galaxy samples (Gil de Paz et al$.$ 2000a,
GIL00a hereafter; Bell \& de Jong 2000; Brinchmann \& Ellis
2000). However, this type of analysis has not yet been performed in a
systematic way, covering a large range of galaxy properties and/or
star formation scenarios. A relevant exception is the work of Ronen,
Arag\'on-Salamanca, \& Lahav (1999) on the Principal Component
Analysis of synthetic galaxy spectra. Similarly, no significant effort
has yet been devoted to determining quantitatively the optimal set of
observables and signal-to-noise ratio required to obtain reliably
derived galaxy properties. Noteworthy exceptions are the works of
Kodama, Bell \& Bower (1999), Bolzonella, Miralles \& Pell\'o (2000),
and Wolf, Meisenheimer \& R\"oser (2001) with regard to the galaxy
classification and redshift determination in broad and medium-band
surveys.

In this paper we explore the uncertainties expected in the derived
properties of galaxies obtained from the analysis of broad-band
photometry of nearby, intermediate, and high-redshift objects. We
quantify these uncertainties using a sample of artificially generated
spectral energy distributions (converted to broadband colors) with
known (input) physical properties, studying the dependence of the
properties derived with the sets of bands available, the observing
errors, as well as the galaxy star formation history and
redshift. This work has interesting predictive capabilities and is
intended to help in the design of future projects on the study of
galaxy evolution. It can be used in the optimization of observing
programs, helping to select the best wavelength bands and
signal-to-noise ratios required to derive precise galaxy
properties. This contribution is mainly focused on the study of the
integrated properties of galaxies, star clusters and H{\sc ii}
regions, although it is our intent to generalize the application to
spatially resolved portions of galaxies as well. A similar approach
has been followed by Charlot \& Longhetti (2001) for the optimization
of emission-line data in galaxy spectra, although no error analysis
was carried out by those authors.

In the Section~\ref{sec2} of this paper we briefly describe the
synthetic galaxy sample. The procedure followed for determining the
uncertainties and degeneracies between the physical properties in this
sample is described in Section~\ref{sec3}. Section~\ref{sec4} includes
a detailed description of the results from this analysis. The main
conclusions are given in Section~\ref{sec5}. Finally, some future
applications for this work are given in Section~\ref{sec6}.

\section {The Sample}
\label{sec2}

We have generated a large number of synthetic galaxy colors
parameterized by different star formation histories, metallicities,
reddening, and redshifts. Because of the unbounded number of possible
combinations we have made some choices and simplifications.

With regard to their star formation history we have considered
galaxies with exponential star formation having timescales ($\tau$)
between 0.2\,Gyr and 6\,Gyr. Although this scenario is a rough
approximation to the actual star formation histories it constitutes
the most widely accepted parameterization of the star formation
history in individual galaxies (see e.g. Brinchmann \& Ellis 2000).

Thus, the properties to be determined are the timescale for the galaxy
formation, the age of the galaxy, the stellar metallicity, and the
reddening. A total of 9000 synthetic galaxies were generated, a
third of them at redshift $z$=0.0, another third at $z$=0.7, and the
remainder at redshift $z$=1.4. These redshift values were chosen to
cover the epoch where most of the evolution of the star formation
activity in the Universe has apparently taken place (Gallego et al$.$
1995; Madau et al$.$ 1996; Connolly et al$.$ 1997; Madau, Dickinson \&
Pozzeti 1998). Since the models available only provide discrete values
in metallicity we chose to assign to each galaxy the nearest
metallicity value of those given by the corresponding model. The
ranges of physical properties covered by the galaxies in the sample
are shown in Table~\ref{table1}.

Although a complete study about the effects of the model uncertainties
on the galaxy properties derived is beyond the scope of this paper,
the impact of the errors in the broad-band colors due to stellar
evolution prescription and spectral calibration uncertainties (Charlot
et al$.$ 1996; Yi, Demarque \& Oemler 1997) will be taken into account
in our further analysis. In addition, in order to illustrate the
effect of these uncertainties we have considered two different sets of
evolutionary synthesis models. The synthetic galaxies were generated
using the predictions for the stellar continuum given by the GISSEL99
models (Bruzual \& Charlot, in preparation) while the best-fitting set
of properties has been derived using both the GISSEL99 and the
P\'{E}GASE models (Version 2.0; see Fioc \& Rocca-Volmerange
1997). Although the theoretical isochrones in both models come mainly
from the Padova group (Bressan et al$.$ 1993), there are fundamental
differences in post-main-sequence evolutionary stages between the two
set of models, including in the early, thermally pulsating, and
post-AGB phases. Noteworthy, the differences in these evolutionary
stages are responsible for the most serious uncertainties in the
stellar population modeling (see Charlot et al$.$ 1996). In
Figure~\ref{fig1} we show the predictions of these two models for
Simple Stellar Population (SSP) and continuous star formation galaxies
with Solar metallicity. For the sake of comparison we have also
included the predictions of the Starburst~99 models (Leitherer et
al$.$ 1999) at ages younger than 0.1\,Gyr.

\section {Analysis}
\label{sec3}

\subsection{Galaxy Colors}
\label{sec3.1}

Once the sample was generated we determined the luminosity (per Solar
mass) in the different bands and the colors for each individual galaxy
in the sample using the predictions of the corresponding evolutionary
synthesis models. We adopted for all the models the same Salpeter IMF
with M$_{\mathrm{low}}$=0.1\,M$_{\odot}$ and
M$_{\mathrm{up}}$=100\,M$_{\odot}$.

The bands considered in this work include the Johnson/Cousins
($UBVR_{C}I_{C}$) and Sloan Digital Sky Survey $u'g'r'i'z'$ (Fukugita
et al$.$ 1996) optical bands, the $JHK$ near-infrared bands, and the
GALEX near-ultraviolet (NUV, 1800-3000\AA) and far-ultraviolet bands
(FUV, 1350-1800\AA; Doliber et al$.$ 2000). Because of the small
differences expected between the standard $K$-band and the
$K_{\mathrm{s}}$ and $K'$ bands, we decided to include only the
standard $K$ in our realizations. In this sense, although the 2MASS
survey has been carried out using $K_{\mathrm{s}}$-band imaging data,
we will refer hereafter to the 2MASS data set when dealing with the
standard $JHK$ bands.

The colors obtained for each stellar population were then reddened
(and the apparent mass-to-light ratios increased) using the
corresponding $E(B-V)$ values and adopting the parameterization of the
Galactic extinction law of Cardelli, Clayton \& Mathis (1989) for a
total-to-selective extinction ratio $R_{V}$=3.1. However, since this
parameterization is only valid down to 1000\AA, and given that some of
the bands selected for this work cover regions of galaxy spectra well
below the Lyman limit (for $z$=0.7 and $z$=1.4), we extended the
extinction law to the far-UV using the A$_{\lambda}$/A$_V$ ratios
given by Mathis (1990; see also Martin \& Rouleau 1989). The
A$_{\lambda}$/A$_V$ mean values for each band (its respective
redshifted rest-frame wavelength) were then computed convolving the
filters response functions with the adopted extinction curve.

Using the number of ionizing Lyman photons predicted by the
evolutionary synthesis models we also computed the contribution of the
nebular continuum and most intense gas emission-lines to all the bands
considered. We assumed that 85\% of the photons with
$\lambda$$<$912\,\AA\ effectively ionize the surronding gas, but a
15~per cent fraction would be observed at the far-ultraviolet or
absorbed by dust (Leitherer et al$.$ 1995; Dove, Shull \& Ferrara
2000; GIL00a).

Most of the 85\% of the far-UV photons absorbed by the surronding gas
in the H{\sc ii} region is reprocessed as nebular continuum and
emission lines in the optical and NIR. However, a very small fraction
of these photons can be re-emitted as free-free radiaton in the
far-UV. In our case, we assumed that these secondary far-UV photons
are not absorbed by neutral gas again, but rather being absorbed by
dust or escaping from the galaxy. In order to determine the nebular
continuum contribution to all the bands we have used the emission and
recombination coefficients given by Ferland (1980) for the near-UV,
optical, and NIR for $T_{\mathrm{e}}$=10$^{4}$\,K. For the
far-UV free-free radiation we have assumed a constant gaunt factor
${\overline g}_{ff}$=1.1 (see Karzas \& Latter 1961) in the range
500-912\AA\ for a gas with $T_{\mathrm{e}}$=10$^{4}$\,K.

With regard to the gas emission-lines we have assumed the relation
between the number of Lyman photons and H$\alpha$ luminosity given by
Brocklehurst (1971) and the theoretical hydrogen line-ratios expected
for a low density gas ($n_{\mathrm{e}}$=10$^{2}$\,cm$^{-3}$) with
$T_{\mathrm{e}}$=10$^{4}$\,K in Case B recombination (Osterbrock
1989). We considered the contribution of the most intense forbidden
lines ([O{\sc ii}]$\lambda\lambda$3726,3729\AA, [O{\sc
iii}]$\lambda\lambda$4959,5007\AA, [N{\sc
ii}]$\lambda\lambda$6548,6583\AA, [S{\sc
ii}]$\lambda\lambda$6717,6731\AA) adopting the mean line ratios
measured by Gallego et al$.$ (1996) for the Universidad Complutense de
Madrid (UCM) sample of local star-forming galaxies (Zamorano et
al$.$ 1994, 1996). Both the nebular continuum and the emission-line
luminosities were corrected for extinction assuming the relation given
by Calzetti, Kinney \& Storchi-Bergmann (1996, see also Calzetti 1997;
Storchi-Bergmann, Calzetti \& Kinney 1994) between the gas and the
stellar continuum reddening associated with the young stellar
population:
$E(B-V)_{\mathrm{stellar}}$=0.44$\times$$E(B-V)_{\mathrm{gas}}$.

Following this procedure we obtain all the input information
concerning the actual properties, colors, and mass-to-light ratios for
the galaxies in the sample. In the next section we discuss our
recovery of these input properties starting from the observed galaxy
colors, and their corresponding measuring errors.

\subsection{Sets of Colors}
\label{sec3.2}

Grouping the colors deduced for these galaxies in different sets and
comparing them with the evolutionary synthesis models allows us to
explore systematics and determine the set of observables that result
in minimum differences between the actual (input) galaxy properties
and the derived (output) properties. This comparison is done using a
combination of Monte Carlo simulations, a maximum likelihood
estimator, and a Principal Component Analysis algorithm.

The number of different combinations of colors that could be
constructed considering a total of 10 potential bands from the UV to
the near-infrared is
${\sum_{r=2}^{10}}{{10!}\over{r!(10-r)!}}=1013$. For this study we
have selected only the 10 sets shown in Table~\ref{table2}. The
detailed comparison of the results obtained for all these sets
provides enough information about the relevance of the different
bands, observing errors, etc$.$, for a precise determination of the
galaxy properties.

\subsection{Comparison Procedure}
\label{sec3.3}

Once the colors of the galaxy sample had been obtained and grouped in
sets we then perturbed the ``observed'' magnitudes by applying random
observing errors. In order to simplify the problem we studied three
cases, corresponding to three different 1-sigma errors in the colors,
0.03, 0.07, and 0.10\,mag, and consider only the case where these
errors are the same in all the colors.

In order to compute the effects of these observing errors in the
galaxy properties to be derived we used a Gaussian distribution of
errors for all bands generated using a Monte Carlo simulation
method. The colors derived for each test-particle were then compared
with the evolutionary synthesis models using a maximum likelihood
estimator, ${\cal L}$. The expression for this estimator (see e.g$.$
Abraham et al$.$ 1999) is

\begin{equation}
{\cal L} = \prod_{n=1}^{N} {1\over \sqrt{2\pi}\Delta C_n}
 \exp\left( - {(c_n-C_n)^2\over 2 \Delta C_n\,^2}\right)
\end{equation}

\noindent
where $C_n$ are the colors derived, $c_n$ are those predicted by the
evolutionary synthesis models and $N$ is the total number of colors
available within each set. Because the same level of error was assumed
for all the bands, the maximization of this expression is equivalent
to computing the minimum $\chi^{2}$. Therefore, in this case, we could
estimate the confidence levels in the galaxy properties {\it via} the
Avni's approximation (Avni 1976) as has been done by Bolzonella et
al$.$ (2000) instead of using numerical simulations. However, the
considerations that lead to this estimation procedure apply only
asymptotically, being applicable when the $\chi^{2}$ estimator
covariance matrix can be replaced by its linear approximation in the
vicinity of the best-fitting set of parameters. Although these
conditions could be fulfilled in our case (Perez-Gonz\'alez et al$.$
2001, in preparation), we have decided to use numerical simulations in
order to be able to derive the degeneracies between the different
galaxy properties.

The ranges in the evolutionary synthesis models parameters where the
data-model comparison was performed are shown in
Table~\ref{table1}. In order to avoid introducing a constraint bias in
the derived properties, the ranges for these comparison were chosen to
be significantly wider than those where the galaxy sample was
generated.

Once the expression ${\cal L}$ is maximized for a significant number
of Monte Carlo test-particles (we used a total of 200) we obtained the
distribution of physical properties associated with the probability
distribution of the galaxy colors. In Figures~\ref{fig2}a, b \& c we
show the results obtained for a nearby galaxy with a exponential star
formation history with $\tau$=4.5\,Gyr timescale, Solar metallicity,
an age of 5.0\,Gyr, and $E(B-V)$ of 0.08\,mag, for observing errors in
the colors of 0.10, 0.07, and 0.03\,mag, respectively. In this case
the set of colors used was that including the GALEX, SDSS, and 2MASS
bands (see Table~\ref{table2}).

These figures illustrate the strong degeneracy between the different
galaxy properties even for relatively small measurement errors. We
have therefore performed a quantitative analysis of these degeneracies
using a Principal Component Analysis (PCA hereafter) on the space of
galaxy properties by solving the eigenvector equation on the
test-particle correlation matrix of each galaxy in the sample (see
Morrison 1976). This analysis gives the direction in the space of
galaxy properties along which the 200 solutions obtained for each
individual galaxy are mainly oriented, constituting the best estimator
of the degeneracy between these properties.

Summarizing, this procedure provides us with (1) the mean derived
properties, (2) the 1-sigma errors, (3) the orientation in the space
of solutions of the Principal Component (PCA1 hereafter), and (4) the
input (i.e$.$ actual) properties of all galaxies in the sample.

\section{Results}
\label{sec4}

Once these quantities had been derived we computed the mean
differences between the output and input values along with the mean
1-sigma spread, at fixed intervals in the input properties. The bins
used were 0.5\,Gyr, 0.025\,dex, and 0.025\,mag in the formation
timescale, age, and reddening, respectively. Mean differences and
1-sigma values in the stellar metallicity were computed for each of
the input values considered.

In Figures~\ref{fig3}a \& b we show the results obtained before and
after computing the mean differences and 1-sigma errors for a
subsample of 500 nearby galaxies with errors in the colors of
0.07\,mag and U+BVRI+JHK data available. Due to the relevance of the
$K$-band luminosities in order to derive stellar masses in nearby
(Arag\'on-Salamanca et al$.$ 1993; GIL00a) and intermediate-redshift
galaxies (see Brinchmann \& Ellis 2000 and references therein), the
mean differences between the derived and input $K$-band mass-to-light
ratios were also computed. Mean uncertainties for all the sets of
observables, redshifts, and observing errors considered are summarized
in Table~\ref{table3}.

In addition, we studied the degeneracies between the galaxy physical
properties analyzing the distribution of the unitary PCA1 vector
components. In Figure~\ref{fig4}a we show the frequency histograms
obtained for the sample of high-redshift galaxies assuming an error in
the colors of 0.07\,mag and the SDSS, SDSS+2MASS, and GALEX+SDSS+2MASS
sets. Note that the PCA1 vector points toward the direction where the
largest fraction of the galaxy properties' variance occurs. In this
sense, a PCA1 vector with components
($u_{\mathrm{log}~t}$,$u_{E(B-V)}$,$u_{\mathrm{log}~Z/Z_{\odot}}$,$u_{\tau}$)=($+$0.707,$-$0.707,0,0),
say, implies the existence of a degeneracy between age and reddening
in the sense that younger, obscured stellar populations have colors
that are indistinguishable from older but less extincted ones. In this
case, no age-metallicity or age-timescale degeneracies would be
present. However, the behavior described above could also result in a
PCA1 vector with components
($u_{\mathrm{log}~t}$,$u_{E(B-V)}$,$u_{\mathrm{log}~Z/Z_{\odot}}$,$u_{\tau}$)=($-$0.707,$+$0.707,0,0). That,
however, would appear in a different position at the frequency
histogram shown in Figure~\ref{fig4}a. Thus, in order to reduce this
sign ambiguity when interpreting our results we forced the
$u_{\mathrm{log}~t}$ component to be positive, changing the sign of
all the vector components if $u_{\mathrm{log}~t}$ was
negative. Finally, in order to quantitatively determine the dominant
degeneracy for each individual galaxy in our sample we have defined
the angle $\theta_{i,j}$ like that satisfying
\begin{equation}
\cos \theta_{i,j} = SIGN\left({u_{i}\over u_{j}}\right) \sqrt{u^{2}_{i}+u^{2}_{j}}
\end{equation}
\noindent
where $u_{i}$ and $u_{j}$ are the $i$ and $j$ components of the PCA1
vector. The angle $\theta_{i,j}$ simultaneously provides a measure of
the angle between the PCA1 vector and the plane of physical properties
$i,j$ and the sign of the degeneracy between the $i$ and $j$
properties. Thus, if $|\cos\theta_{i,j}|$$\simeq$1 the degeneracy
between the $i$ and $j$ properties would dominate the total
degeneracy. Moreover, if $\cos\theta_{i,j}$$>$0 an increase in both
the $i$ and $j$ properties would lead to similar observational
properties, while if $\cos\theta_{i,j}$$<$0 the value of one of the
properties should decrease. In Figure~\ref{fig4}b we show the
distribution of $\cos\theta_{i,j}$ as function of the age for the
high-redshift sample assuming an error in the colors of 0.07\,mag and
the GALEX+SDSS+2MASS set available.

Along this section we will describe the results obtained from the
analysis of the distributions shown in Figures~\ref{fig3} \&
\ref{fig4} for the different redshifts, observing errors, and
band-pass combinations considered.

\subsection{Nearby galaxies}
\label{sec4.1}

\subsubsection{Formation Timescale} 
\label{sec4.1.1}

With regard to the formation timescale in nearby galaxies,
Figure~\ref{fig5}a indicates that, even for relatively small observing
errors, its uncertainty is very high (see also Table~\ref{table3}) and
shows a strong dependence with the value of the formation timescale
itself. The larger uncertainty in the formation timescale for larger
values of this quantity is mainly due to the small sensitivity of the
optical-NIR colors of stellar populations with ages $t$$<<$$\tau$ to
changes in its formation timescale. The use of $U$-band data
significantly reduces this uncertainty, probably due to the high
sensitivity of this band to the presence of recent star formation that
allows to rule out instantaneous-burst solutions when recent star
formation associated with larger $\tau$ values has effectively taken
place. The use of NIR data, however, does not provide relevant
information about the formation timescale of the stellar
population. Moreover, the reduction achieved in the uncertainties of
the different galaxy properties by using $JHK$ data compared with
those obtained using exclusively $K$-band data is very small (see
Table~\ref{table3}). As we will show in Sections~\ref{sec4.2} \&
\ref{sec4.3} this is not the case for the intermediate and
high-redshift galaxies, where these bands now cover the redshifted
optical spectrum. Finally, in the same way that the $U$-band, the use
of UV data provides an additional reduction in the formation
timescale. As it is clearly seen in Figure~\ref{fig2}, for a particular
galaxy the formation timescale is mainly degenerate with the age of
the stellar population, in the sense that, within the observing errors
assumed, an increase in the formation timescale accompanied by an
increase in the age can result in similar UV-optical-NIR colors. Due
to this age-timescale degeneracy part of the reduction in the
timescale uncertainty obtained by the use of UV data can be explained
by the significant reduction in the age uncertainty achieved by
including UV data (see below).

\subsubsection{Age} 
\label{sec4.1.2}

With respect to the age determination in nearby galaxies ($z$=0), the
Figure~\ref{fig5}a also shows that a significant reduction in the age
uncertainty is achieved by including NIR data. It is important to note
that the use of additional NIR data result in the same dominant
degeneracy that if only optical data are used (see below), but the
range of physical properties where this degeneracy takes place is
significantly smaller. The most significant improvement in the age
determination, however, is obtained when UV data are available. This
is mainly due to the emission arising from post-AGB stars in
low-metallicity populations and at ages younger than 10\,Gyr and to
the ``UV-upturn'' in high-metallicity evolved ($t$$>$10\,Gyr) stellar
populations that result in highly peculiar UV-optical colors. However,
the uncertainty in the modeling of post-AGB stars (Charlot et a$.$
1996) and the low-mass core helium-burning Horizontal Branch (HB
hereafter) and evolved HB stars that lead to the ``UV-upturn'' (Yi et
al$.$ 1997), introduce additional errors in the UV-optical colors
during the data-models comparison. Charlot el al$.$ (1996) estimated
using two different theoretical prescriptions that the uncertainty
only in the post-AGB phase modeling could result in differences of
about 1\,mag in the UV-optical colors of a several-Gyr-old stellar
population. Moreover, although we assumed the same observing errors
for all the bands, the faint UV emission of evolved stellar
populations is expected to result in very large observing errors in
the UV-optical colors. Therefore, while the stellar evolution of these
stars is not well understood the age determination in old stellar
populations should not rely on the use of UV data.

Along with the formation timescale, the age of the stellar population
in nearby galaxies is mainly degenerate with the dust extinction, in
the sense that older stellar populations with low dust content have
similar colors to highly-extincted, younger stellar
populations. Although in the case of very old stellar populations the
age-extinction degeneracy also competes with the age-metallicity
degeneracy (see Worthey 1994), the age-extinction degeneracy is still
dominant in this range for all the band-pass combinations and
observing errors considered in this work. It could be argued that the
strong discretization of the metallicity in our models could be
responsible for the relatively weak age-metallicity degeneration
derived. However, the fact that this behavior is observed even for the
largest observing errors considered indicates that it is real and a
natural consequence of the use of broad-band data. In this sense, the
combination of broad-band with narrow- or medium-band data or
spectroscopic indexes would break the age-extinction
degeneracy, making of the age-metallicity the dominant degeneracy (see
Worthey 1994).

\subsubsection{Dust Extinction} 
\label{sec4.1.3}

The dust extinction is derived with a very high accuracy
($E(B-V)$=0.04-0.20\,mag) even for large observing errors and
relatively low number of observables (see Table~\ref{table3}). In the
case of the nearby galaxies, the uncertainty in the dust extinction
does not depend on the value of the dust extinction itself and is
mainly degenerate with the age of the stellar population (see above)
with some contribution from the extinction-metallicity degeneracy. In
combination with $BVRI$ optical data either the use of UV, $U$, or NIR
data provide a significant reduction in the dust-extinction
uncertainty. In order to better derive the dust extinction the use of
a wider wavelength baseline in wavelength (e.g$.$ using UVIJK) is more
effective than fully covering the optical range (UBVRI). This is
mainly due to the reduction in the metallicity uncertainty by the use
of NIR data (see below) that leads, via the extinction-metallicity
degeneracy, to a reduction in the dust-extinction uncertainty. Again,
the use of $JHK$ data instead of only $K$-band data do not lead to a
significant reduction in the dust-extinction uncertainties.

\subsubsection{Metallicity}
\label{sec4.1.4}

With regard to the metallicity of the stellar population the
uncertainties derived are strongly dependent on the band-pass
combination available and the value of the metallicity itself. In
particular, the uncertainties derived are smaller as the metallicity
becomes higher (see Figure~\ref{fig5}a). Within the age range
considered, the main contributors to the optical and NIR emission of
SSP galaxy are the main-sequence and RGB stars. However, for a more
constant star formation, a significant contribution from
core-Helium-burning stars is expected (see Charlot \& Bruzual
1991). In order to determine the source of the metallicity dependence
of these uncertainties we have produced the same diagrams shown in
Figure~\ref{fig5}a but restricted to formation timescales shorter than
50\,Myr. The analysis of this diagram shows no dependence of the
uncertainties with metallicity, which implies that the source of the
dependence was the distinct photometric evolution of high-metallicity
core Helium burning stars (Mowlavi et al$.$ 1998). It is worth noting,
however, that at very high metallicities the uncertainties in the
modeling of the stellar populations are themselves very large because
of the lack of very metal rich stars of any age in the Solar
neighborhood that could be used as spectral calibrators (see Charlot
et al$.$ 1996).

The most significant reduction in the mean metallicity uncertainty is
achieved when NIR data are used in combination with optical data (see
Figure~\ref{fig5}a). Although the uncertainties in the model
predictions for the thermally pulsating AGB (TP-AGB hereafter) and the
upper RGB can result in differences in the ($V-K$) color predicted by
different models of $\sim$0.10-0.15\,mag (Charlot et al$.$ 1996), the
improvement in the metallicity determination by the use of NIR data is
still relevant. In this sense, in Table~\ref{table3} we show that the
mean metallicity uncertainty for the U+BVRI set is 0.32\,dex assuming
an observing error of 0.03\,mag, while the uncertainty for the
U+BVRI+K set assuming an observing error of 0.10\,mag is only
0.26\,dex.

\subsubsection{Stellar Mass} 
\label{sec4.1.5}

As input for the $K$-band mass-to-light ratio of the stellar
populations we have adopted M$_{K,\odot}$=3.33 (Worthey 1994). It
should be noticed that along with the errors in the stellar
mass-to-light ratios derived here the misunderstanding of the actual
IMF introduce an additional, systematic uncertainty, which, in fact,
constitutes the most important source of error in the determination of
the galaxy stellar mass (Bell \& de Jong 2001). In addition, the poor
constraints on the theoretical isochrones of upper-RGB stars and AGB
stars can result in a 20~per cent uncertainty in the $K$-band
mass-to-light ratio (Charlot et al$.$ 1996). In Figures~\ref{fig6}a \&
\ref{fig6}b we show the uncertainties expected in the $K$-band
mass-to-light ratio from different sets of observables that include
$K$-band data. These uncertainties show a strong dependence with the
galaxy age and formation timescale in the sense that larger
uncertainties are expected at lower values of the formation timescale
and older ages. Figure~\ref{fig5}a shows that the value of the age
uncertainty (in log~$t$ scale) is almost independent of the age
itself. In addition, Figure~\ref{fig1} indicates that the rate of
change in the $K$-band mass-to-light ratio (with log~$t$) is higher
when the stellar population becomes older, specially for very low
values of the formation timescale. Therefore, for a constant
uncertainty in log~$t$, an increase in the uncertainty of the
mass-to-light ratio at very old ages is expected.

Figure~\ref{fig6}a also shows that the mass-to-light ratio
determination is biased toward lower values. This bias, which is
particularly important at old ages, is probably due to the upper limit
of 15\,Gyr in age imposed during the data-models comparison (see
Table~\ref{table1}), although other contributors can not be ruled out
(see Sections~\ref{sec4.2.2} \& \ref{sec4.3.2}). As it is clearly seen
in Figure~\ref{fig6}b, the use of UV data allows to reduce both the
uncertainty and bias in the mass-to-light determination. This
reduction is directly related with the reduction in the age
uncertainty described above. However, as we already commented, the use
of UV data for the study of stellar populations with ages older than
several Gyr can lead to wrong conclusions because of the uncertain
modeling of the post-AGB phase and the ``UV-upturn''.

The behavior described above for the timescale, age, dust extinction,
metallicity, and stellar mass is identical for any observing error but
with larger mean uncertainties for larger observing errors. The reader
is referred to the Table~\ref{table3} for the dependence of the mean
uncertainties in the different galaxy properties derived with the
observing errors.

\subsection{Intermediate-redshift galaxies}
\label{sec4.2}

\subsubsection{Formation Timescale} 
\label{sec4.2.1}

In Figure~\ref{fig7} we show the uncertainties derived for the
properties of intermediate-redshift galaxies ($z$=0.7). With regard to
the formation timescale the uncertainties are very large (2-3\,Gyr),
even larger than for the nearby galaxy sample. As we commented in
Section~\ref{sec4.1}, the optical-NIR colors are quite insensitive to
changes in the formation timescale with $t$$<<$$\tau$. Therefore,
since we are assuming that these galaxies are statistically younger
than the those observed in our Local Universe (see Table~\ref{table1})
and the range in formation timescale is obviously the same, the
uncertainty in the formation timescale is necessarily higher. For the
same reason the uncertainty at very low timescale values is much lower
than at high timescale values.

The upper panel of Figure~\ref{fig7}a also suggests a significant bias
in the timescale determination toward lower values of this
property. This bias is also the consequence of the small changes in
the optical-NIR colors of these galaxies with the timescale when the
age is younger than the timescale value. In this case, the higher rate
of change in the colors toward lower formation timescales
systematically leads to lower values in order to reproduce the
probability distribution associated with the observing errors. It is
worth noting that, because of the reduction of this bias, the use of a
larger number of bands may result in some cases in a higher timescale
uncertainty (see Table~\ref{table3} for the results on the U+BVRI+K
and U+BVRI+JHK sets). Like in the nearby galaxies case, the dominant
degeneracy involving the formation timescale is the age-timescale
degeneracy, in the sense that older galaxies with high formation
timescales have similar colors that younger galaxies with a more
instantaneous star formation. This is true for any band-pass
combination considered. With regard to the optimal set of observables,
Table~\ref{table3} demonstrates that for the same number of bands the
use of wider wavelength baselines results in lower uncertainties. In
particular, the use of the UVIJK set reduces the timescale, age, and
metallicity uncertainties inherent to the UBVRI set providing also a
much lower dust-extinction uncertainty than the BVRI+K set. On the
other hand, the SDSS+2MASS and GALEX+SDSS+2MASS sets result in very
similar uncertainties (see Table~\ref{table3}), which implies that the
optical and NIR bands provide most of the information available in the
UV and in the blue part of the optical spectrum about the galaxy age,
star formation history, and metallicity.

\subsubsection{Age} 
\label{sec4.2.2}

With respect to age of the intermediate-redshift galaxies the
uncertainties derived are significantly smaller than in the
nearby-galaxies case. This is mainly due to the higher rate of change
in the rest-frame optical colors within the age range assumed for
these galaxies compared with that assumed for the nearby galaxies (see
Table~\ref{table1} and Figure~\ref{fig1}). In addition, the fact that
the $K$-band now corresponds to the rest-frame $J$-band emission
implies that the effect of the uncertainties in the model predictions
associated with the upper RGB and AGB evolutionary stages is less
important (see Section~\ref{sec4.4.2}). On the other hand, the use of
$U$-band data for determining ages older than 1\,Gyr at these
redshifts is strongly limited by the uncertainty in the modeling of
the rest-frame UV emission from post-AGB stars (Charlot et al$.$ 1996;
see Section~\ref{sec4.1}). However, the most significant decrease in
the age uncertainty is achieved when NIR data are used, specially if
data in all the bands ($JHK$) are available. This is probably due to
the fact that the $JHK$ set provides information simultaneously about
the presence of AGB stars (via the rest-frame $z'$ and $J$ bands) and
main-sequence stars (via the rest-frame $R$-band).

Figure~\ref{fig5}a also shows the existence of a significant bias
toward younger ages for the BVRI and UBVRI sets. In this case the
presence of this bias is due (1) to the existence of a formation
timescale bias and a strong age-timescale degeneracy and (2) to the
fact that the optical colors of the stellar populations change more
slowly as the population becomes older. In the latter case, in order
to reproduce the distribution of optical colors associated with the
observing errors, the best-fitting solution should be found at younger
ages, where the intrinsic dispersion of the model colors is larger. As
we show below a bias in age also results in a bias in the galaxy
$K$-band mass-to-light ratio. The use of wider wavelength baselines
allows to significantly reduce this bias. In particular, the use of
the UVIJK leads to a less severe bias and lower age uncertainties than
the U+BVRI and the U+BVRI+K sets. Within the age uncertainty interval
the degeneracy is mainly dominated by the age-timescale degeneracy
with some contribution from the age-extinction degeneracy in those
band-pass combinations that do not include UV or $U$-band data.

\subsubsection{Dust Extinction} 
\label{sec4.2.3}

The dust extinction in the sample of intermediate-redshift galaxies is
derived with high accuracy, specially when $U$-band data are available
(see Figure~\ref{fig7}a). In this case, the uncertainties expected in
$E(B-V)$ are in any case smaller than 0.10\,mag for observing
uncertainties as high as $\Delta C_n$=0.10\,mag. The significant
reduction achieved, if we compare these results with those derived for
the nearby galaxies, is due to the very high sensitivity of the
redshifted UV emission to the presence of small amounts of dust. In
those band-pass combinations not including $U$-band data we notice a
clear dependence of the dust-extinction uncertainty with the value of
the extinction itself, with larger uncertainties at larger values of
the extinction (see Figure~\ref{fig7}a). The analysis of the PCA1
components also indicates that at dust-extinction values higher than
$E(B-V)$$>$0.5\,mag the age-extinction degeneracy becomes very
important. This implies that in highly extincted intermediate-redshift
galaxies a small increase in the amount of dust can lead to the same
optical colors (specially if $U$-band data are not used) that a
comparable decrease in the age of the stellar population would
produce.

\subsubsection{Metallicity} 
\label{sec4.2.4}

With regard to the metallicity uncertainty, Figure~\ref{fig7}a shows
that the uncertainty decreases with the value of the metallicity
itself. The reduction is particularly important when NIR data are
available. The use of the three $JHK$ NIR bands reduces this
uncertainty over the whole range of metallicities. In this sense, the
use of the UVIJK set results in lower metallicity uncertainties than
the U+BVRI and the U+BVRI+K sets (see Table~\ref{table3}). It is
important to keep in mind that the $JHK$ filters now cover the
rest-frame $R$, $z'$, and $J$ bands. In the age range considered the
main contribution to the rest-frame optical emission comes from
main-sequence stars. On the other hand, the rest-frame NIR emission,
along with main-sequence stars, shows an important contribution from
AGB and core-Helium-burning stars (see Charlot \& Bruzual 1991). The
role played by AGB stars is more relevant if the formation is
instantaneous, while the core-Helium-burning stars may dominate the
total NIR emission for a more constant star formation
scenario. Therefore, the behavior described above is probably due to
the distinct evolution of high-metallicity AGB stars (see Willson 2000
and references therein) and core-Helium-burning stars (Mowlavi et
al$.$ 1998) compared with the relatively well-defined sequence in
their evolutionary properties established for sub-solar
metallicities. Within the error intervals derived, the metallicity is
mainly degenerate with the age, specially in those sets including
$U$-band data. This is probably due to the reduction in the
age-extinction degeneracy thanks to the information provided by the
$U$-band data about the rest-frame UV.

\subsubsection{Stellar Mass} 
\label{sec4.2.5}

The comparison between Figures~\ref{fig6}b and \ref{fig7}b shows that
the mean uncertainties in the $K$-band mass-to-light ratio (or stellar
mass) of intermediate-redshift galaxies are much lower than those
derived for the nearby sample. First, it is important to note that in
these figures we represent absolute errors. For a Solar-abundant
12\,Gyr-old nearby galaxy formed instantaneously the $K$-band
mass-to-light ratio is $\sim$1.3\,M$_{\odot}$/L$_{K,\odot}$, while for
a 5\,Gyr-old galaxy at $z$=0.7 is
$\sim$0.4\,M$_{\odot}$/L$_{K,\odot}$. Therefore, the relative
uncertainties, assuming the average absolute uncertainties given in
Table~\ref{table3} for $\Delta C_n$=0.07\,mag, would be about 30 and
20~per cent, respectively for the nearby and intermediate-redshift
galaxies. Although this still implies a significant improvement in the
$K$-band mass-to-light ratio determination, it is also noticeable that
the $K$ filter now traces the rest-frame $J$-band luminosity, which is
more affected by the misunderstanding about the actual IMF (see Bell
\& de Jong 2001). Finally, the $J$-band luminosity is also more
sensitive to small differences between the assumed exponential star
formation and the galaxy actual star formation history than the
rest-frame $K$-band data.

As we pointed out in Section~\ref{sec4.2.2} the bias in the age
determination toward lower age values also leads to a strong bias in
the $K$-band mass-to-light ratio of intermediate-redshift galaxies due
to the systemic decrease in the rest-frame $J$-band luminosity per
Solar mass with the age of stellar population when the age is older
than $\sim$1\,Gyr.

\subsection{High-redshift galaxies}
\label{sec4.3}

\subsubsection{Formation Timescale} 
\label{sec4.3.1}

With regard to the formation timescale, Figure~\ref{fig8}a shows that
the bias toward lower timescale values observed at intermediate
redshift is even more pronounced at high-redshift. This bias is a
natural consequence of the difficulty of deriving/predicting the
long-term star formation history of a galaxy when it is still very
young. This is also evidenced by the fact that the mean timescale
uncertainty increases systematically with redshift for the same
observing errors and band-pass combinations. In Table~\ref{table3} we
also show that in many cases (BVRI vs$.$ U+BVRI; SDSS vs$.$
SDSS+2MASS) the mean timescale uncertainties increase when a larger
number of observing bands is used, with a progressive reduction in
this bias. As in the intermediate-redshift case the dominant
degeneracy involving the galaxy formation timescale occurs with the
age of the stellar population.

\subsubsection{Age} 
\label{sec4.3.2}

The large formation timescale uncertainty described above and the
existence of a strong age-timescale degeneracy, specially at ages
older than 100\,Myr, lead to very large age uncertainties, even larger
than those derived for the intermediate-redshift galaxies.  The
age-timescale degeneracy at ages younger than 100\,Myr is
significantly smaller because at these young ages a change in the
formation timescale, which ranges between 200\,Myr-6\,Gyr (see
Table~\ref{table1}), does not affect to the UV-optical-NIR colors of
the stellar population. In other words, the degeneracy in timescale
within this age range is complete and no correlation between the age
uncertainty and any other uncertainty is expected. In this case the
main degeneracies are the age-extinction and the age-metallicity ones.

Moreover, the age-timescale degeneracy in combination with the bias in
formation timescale described above are also responsible for the
strong bias in age observed in Figure~\ref{fig8}a at ages older than
$\sim$50\,Myr. The fact that the UVIJK set provides a better age and
timescale determination than the U+BVRI and U+BVRI+K sets demonstrates
the importance of obtaining $JHK$ data in order to derive the
properties of high-redshift galaxies. This is due to the fact that the
$JHK$ filters now cover the rest-frame $V$, $R$, and $z'$ optical
bands, where the changes due to galaxy evolution are more noteworthy
and the information content about the galaxy properties is larger. In
particular, the $JHK$ filters would provide information simultaneously
about the presence of main-sequence stars (via the rest-frame $V$ and
$R$ bands), core-Helium-burning stars (via the rest-frame $V$, $R$,
and $z'$ bands), and AGB stars (via the rest-frame $z'$-band;
$t$$>$0.5\,Gyr).

\subsubsection{Dust Extinction} 
\label{sec4.3.3}

Because of the extensive coverage of the UV range of the spectrum, the
study of high-redshift galaxies using optical-NIR colors leads to
very small dust-extinction uncertainties. In this sense, the
dust-extinction uncertainties given in Table~\ref{table3} at this
redshift assuming an observing error of 0.10\,mag are in the range
$E(B-V)$=0.03-0.07\,mag. The dust extinction within the interval of
uncertainty is mainly degenerate with the age of the stellar
population.

\subsubsection{Metallicity} 
\label{sec4.3.4}

Figure~\ref{fig8} shows that the metallicity uncertainty for the
high-redshift sample does not show the strong metallicity dependence
found in nearby and intermediate-redshift samples. Only when $JHK$ NIR
data are available the uncertainties at very high metallicities become
significantly smaller than those derived for the low metallicity
galaxies. As we commented in Section~\ref{sec4.2.4} for the
intermediate-redshift case, this is probably due to the distinct
signature of high-metallicity core-Helium-burning stars (eg$.$ in the
number ratio of blue-to-red supergiants; Mowlavi et al$.$ 1998) within
the age range considered. In the case of a SSP galaxy, these stars
dominate the rest-frame $VRz'$ ($JHK$ at $z$=1.4) emission for ages
younger than 0.4\,Gyr, while the emission at shorter wavelengths comes
mainly from main-sequence stars (see Charlot \& Bruzual 1991). It is
important to note that the core-Helium-burning stars may dominate the
emission in the $R$ and $z'$ bands up to ages of 5\,Gyr for larger
formation timescales. Within the uncertainty intervals obtained, the
metallicity is mainly degenerate with the age of the stellar
population.

\subsubsection{Stellar Mass} 
\label{sec4.3.5}

The $K$-band mass-to-light ratio uncertainties derived here are very
small compared with those obtained from the nearby and
intermediate-redshift samples, with values ranging between 0.01 and
0.06\,M$_{\odot}$/L$_{K,\odot}$. If we adopt a $K$-band mass-to-light
of 0.27\,M$_{\odot}$/L$_{K,\odot}$, which corresponds to the value
expected for a 3\,Gyr-old galaxy with Solar metallicity, the relative
uncertainty would range between 5\% and 20\%, depending of the
band-pass combination available. Figure~\ref{fig8}b shows that there
is also a strong dependence of the mass-to-light ratio uncertainty
with the value of the mass-to-light ratio itself. In particular, a
clear minimum in its uncertainty is observed at ages older than
8\,Myr, which is probably associated with the evolution of the massive
stars off the main sequence toward the red supergiant phase. During
this part of the evolution a sudden change in the rest-frame $z'$
luminosity and optical colors of a SSP is produced, which could
explain why the uncertainty is particularly small around this age
value.

\subsection{Effects of the Model Uncertainties}
\label{sec4.4}

In this section we analyze the results obtained when the optical-NIR
colors of a sample of galaxies generated using the GISSEL99 models are
compared with the predictions of the P\'{E}GASE evolutionary synthesis
models. We have restricted this comparison to the nearby sample and
the range of properties specified in Table~\ref{table1}. The results
of this comparison are shown in Figure~\ref{fig9}.

\subsubsection{Formation Timescale} 
\label{sec4.4.1}

Figure~\ref{fig9}a shows that the same bias toward lower values of the
formation timescale that we noted for the intermediate and
high-redshift samples is also present in this case (see
Sections~\ref{sec4.2.1} \& \ref{sec4.3.1}). The main reason for the
existence of this bias is the small change in the optical-NIR colors
of the stellar population with the timescale when the age
$t$$<<$$\tau$. Therefore, in order to compensate both the observing
errors and the differences in the color predictions between the
GISSEL99 and P\'{E}GASE models, the best-fitting solution has to be
found at lower values of the timescale where the intrinsic dispersion
of the colors is larger. The existence of this strong bias also leads
to very small timescale uncertainties compared with those obtained
using the GISSEL99 models. Within the uncertainty intervals derived,
the dominant degeneracy involving the galaxy formation timescale is
the age-timescale degeneracy.

\subsubsection{Age} 
\label{sec4.4.2}

\begin{figure}
\caption{Comparison of the predictions of the Bruzual \& Charlot (in
preparation; {\it thick-lines}), P\'{E}GASE (version 2.0; see Fioc \&
Rocca-Volmerange 1997; {\it thick grey lines}), and Starburst~99
(Leitherer et al$.$ 1999; {\it thin-lines}) models for a Solar
metallicity galaxy with instantaneous (SSP; panel {\bf a}) and
continuous star formation (panel {\bf b}).
\label{fig1}}
\resizebox{0.95\hsize}{!}{\includegraphics*[131,293][482,495]{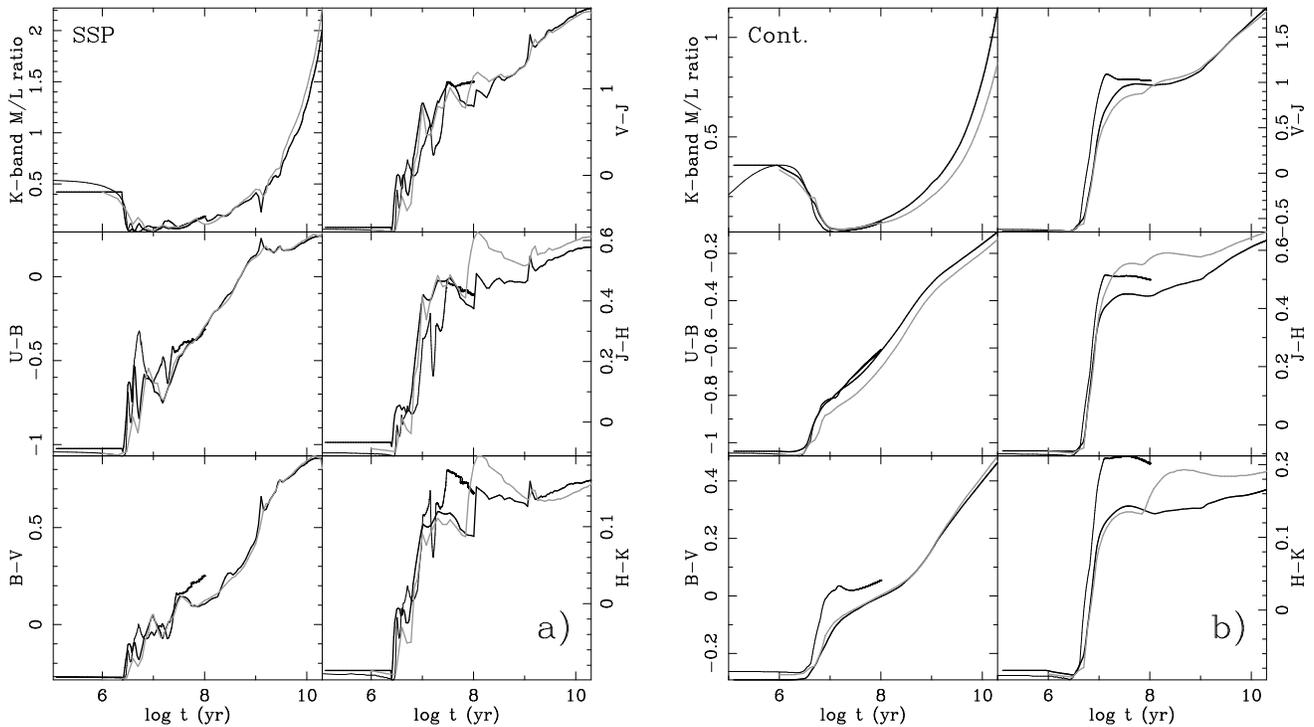}}
\end{figure}

\begin{figure}
\caption{{\bf a, b \& c)} {\it Top panels}: Distribution of solutions
associated with the 200 Monte Carlo simulations in the {\it
timescale-age} space for a nearby galaxy with a formation timescale of
4.5\,Gyr, an age of 5\,Gyr, a dust extinction $E(B-V)$=0.08\,mag, and
Solar metallicity. {\it Middle panels}: Distribution of solutions in
the {\it age-metallicity} space for the same galaxy model. {\it Bottom
panels}: The same for the {\it age-extinction} space. Observing errors
are 0.10, 0.07, and 0.03\,mag, respectively for the {\bf a}, {\bf b},
and {\bf c} figures. The input properties of the galaxy are marked
with a star symbol. This comparison was performed for the
GALEX+SDSS+2MASS color set. The size of each point is proportional to
the value of the maximum likelihood estimator, ${\cal L}$, for the
corresponding Monte Carlo simulation. Degeneracies between the galaxy
properties are evidenced by correlations in the distribution of
points.
\label{fig2}}
\resizebox{0.9\hsize}{!}{\includegraphics*[133,212][486,571]{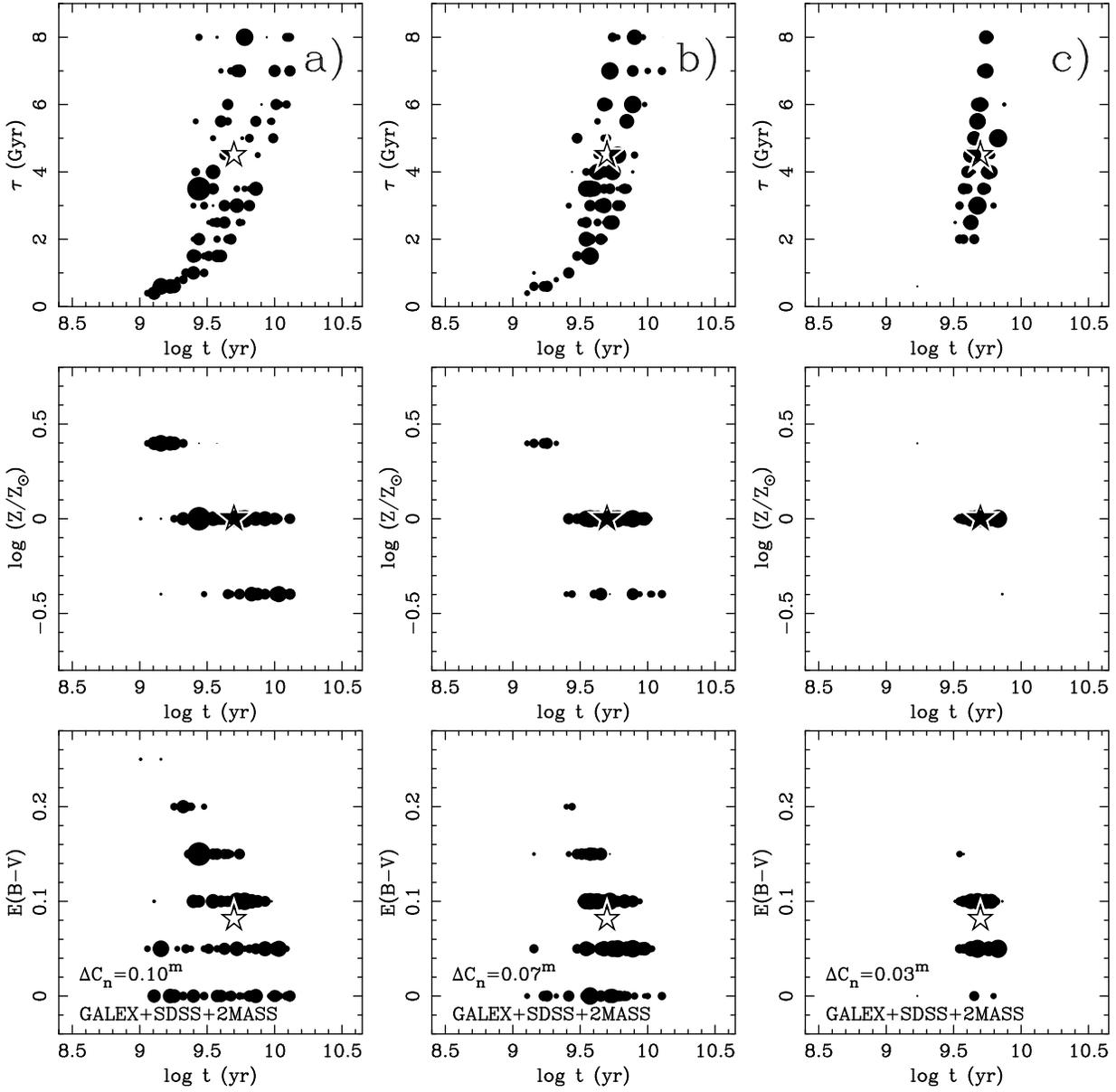}}
\end{figure}

\begin{figure}
\caption{Uncertainties in the derived properties as function of the
properties themselves for a subsample of 500 nearby galaxies and the
U+BVRI+JHK set of observables. In Panel {\bf a} the mean differences
between the derived and the input properties computed in intervals of
0.5\,Gyr in formation timescale, 0.025\,mag in color excess, and
0.05\,dex in age are represented by a light-grey line. Mean
$\pm$1\,$\sigma$ values for the derived properties are delimited by
two dark-grey lines. The region defined by the
mean\,$\pm$\,1\,$\sigma$ lines is also represented in Panel {\bf
b}. Grey shaded areas represent the regions not covered by our
comparison procedure. 
\label{fig3}}
\resizebox{0.85\hsize}{!}{\includegraphics*[149,238][485,548]{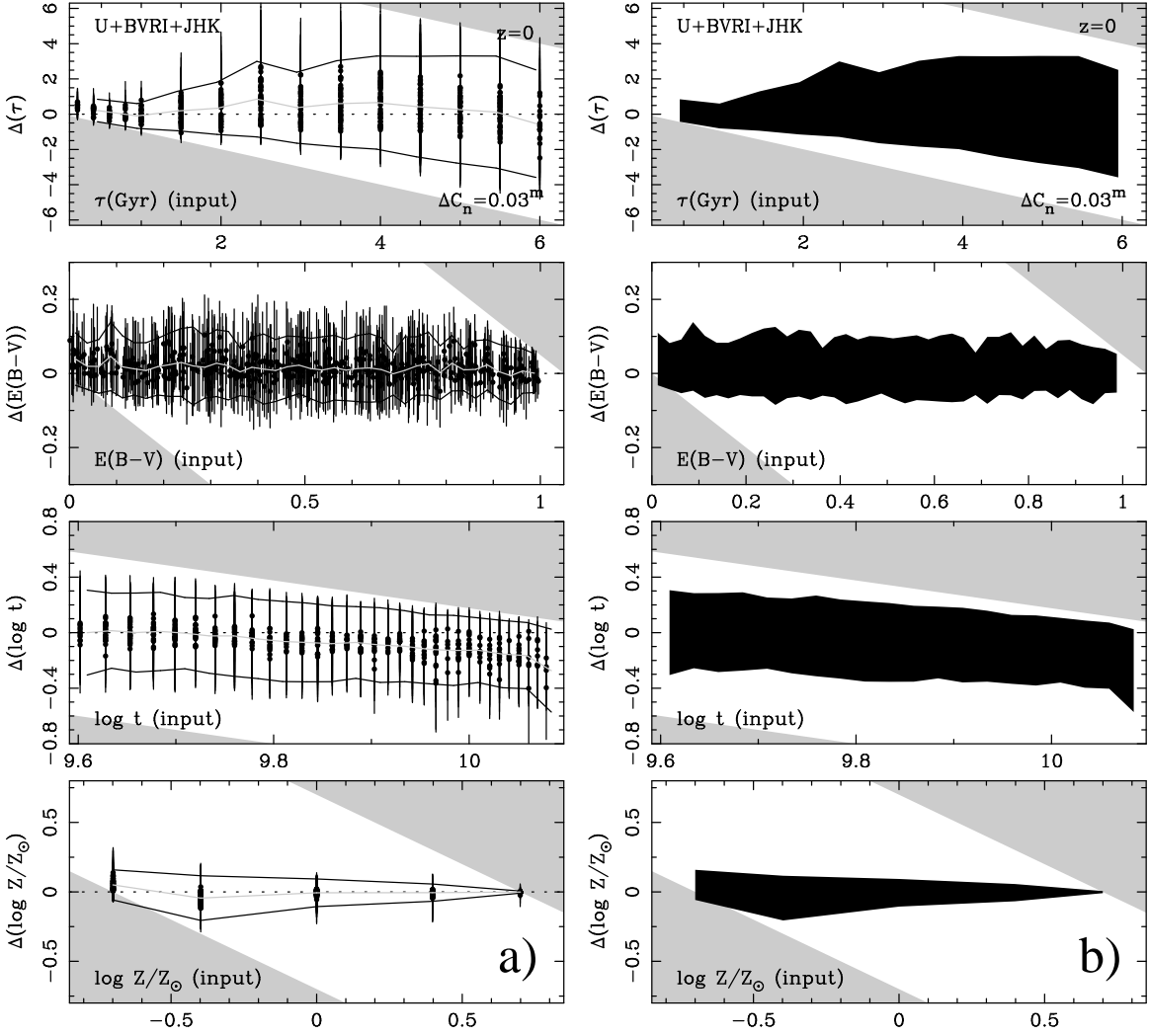}}
\end{figure}

\begin{figure}
\caption{{\bf a)} Frequency histograms for the components of the PCA1
vectors for the sample of high-redshift galaxies with $\Delta
C_n$=0.07\,mag for three different sets of observables (SDSS,
SDSS+2MASS, and GALEX+SDSS+2MASS). This figure suggests that the
age-metallicity and age-extinction degeneracies are comparable for the
SDSS set, while the age-extinction and the age-timescale degeneracies
are competing in the case of the SDSS+2MASS and GALEX+SDSS+2MASS
sets. {\bf b)} Cosine of the angle between the PCA1 vector and the
age-timescale ({\it top}), age-extinction ({\it middle}), and
age-metallicity ({\it bottom}) planes, as measure of the degeneracy
between the stellar populations properties (see Section~\ref{sec4}),
as funcion of the age for high-redshift galaxies with an uncertainty
$\Delta C_n$=0.07\,mag and the GALEX+SDSS+2MASS set. The sign of the
cosine indicates if the degeneracy is in the sense that an increase in
both properties can lead to the same observational properties
(positive) or the value for one magnitude has to be decreased while
the other is increased (negative). From this figure is clear the
dependence of the dominant degeneracy with the age of the stellar
population, going from the age-metallicity and age-extinction
degeneracies at ages below 10\,Myr, to only the age-extinction at ages
between 10\,Myr and 300\,Myr, and the age-timescale at ages between
300\,Myr and 3\,Gyr (see Section~\ref{sec4.3.2}). \label{fig4}}
\resizebox{0.8\hsize}{!}{\includegraphics*[128,242][488,532]{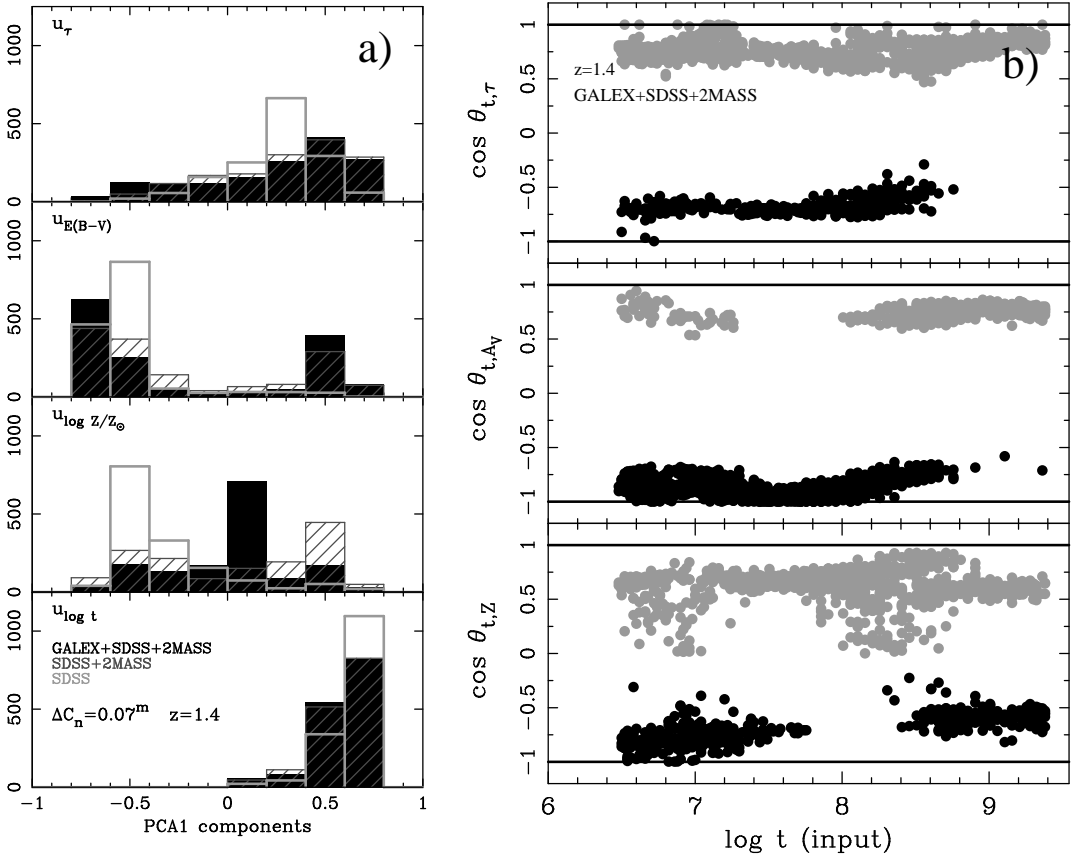}}
\end{figure}

\begin{figure}
\caption{Uncertainties in the derived properties of nearby
galaxies for the BVRI, U+BVRI, and U+BVRI+K sets (panel {\bf a}) and
the SDSS, SDSS+2MASS, and GALEX+SDSS+2MASS sets (panel {\bf b})
assuming observing errors of 0.03\,mag.\label{fig5}}
\resizebox{0.93\hsize}{!}{\includegraphics*[133,235][485,556]{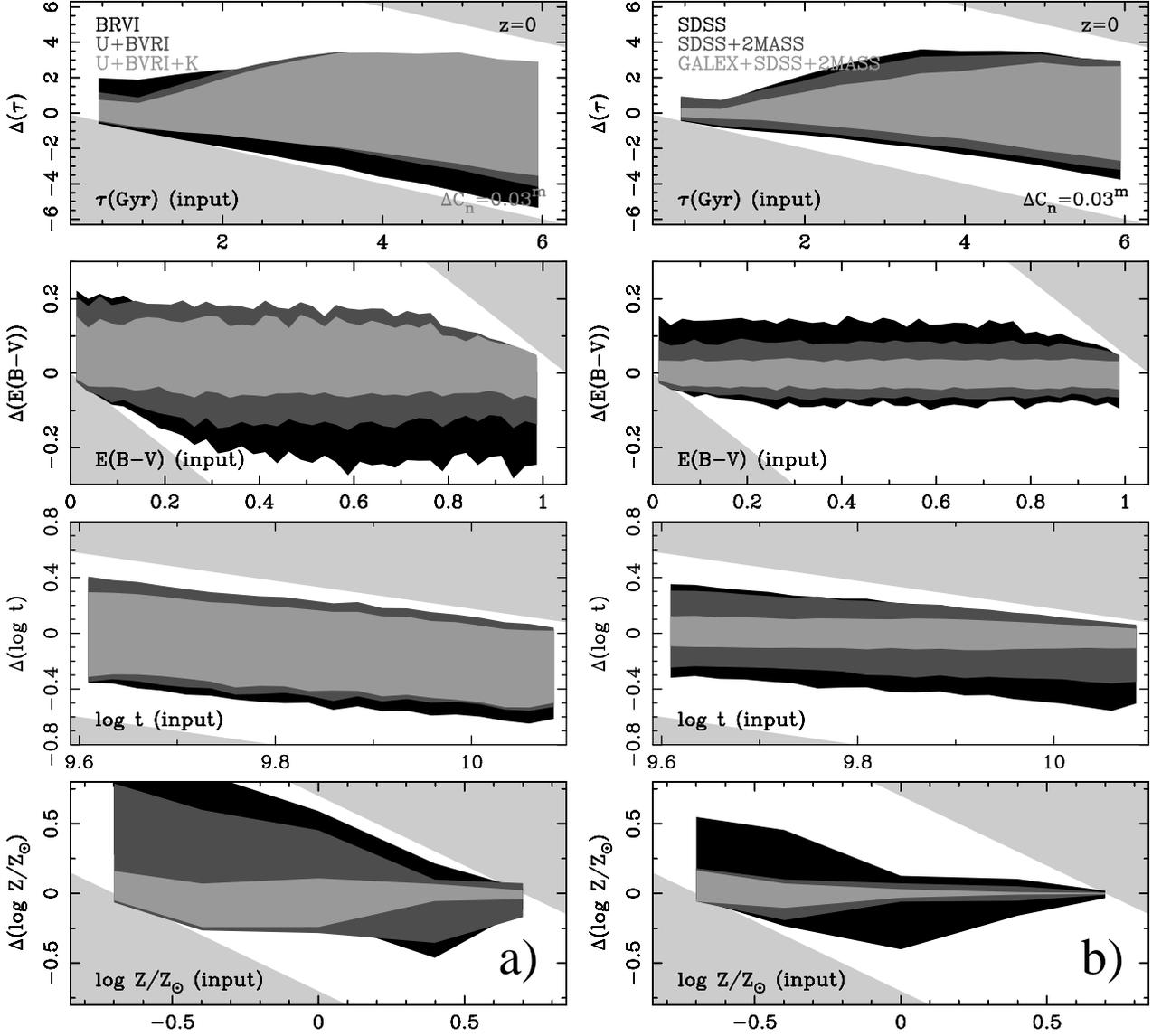}}
\end{figure}

\begin{figure}
\caption{Uncertainties in the derived $K$-band
mass-to-light ratios for the BVRI+K, U+BVRI+K, and U+BVRI+JHK sets
(panel {\bf a}) and the BVRI+K, SDSS+2MASS, and GALEX+SDSS+2MASS sets
(panel {\bf b}) assuming observing errors of 0.03\,mag. \label{fig6}}
\resizebox{0.93\hsize}{!}{\includegraphics*[133,235][485,556]{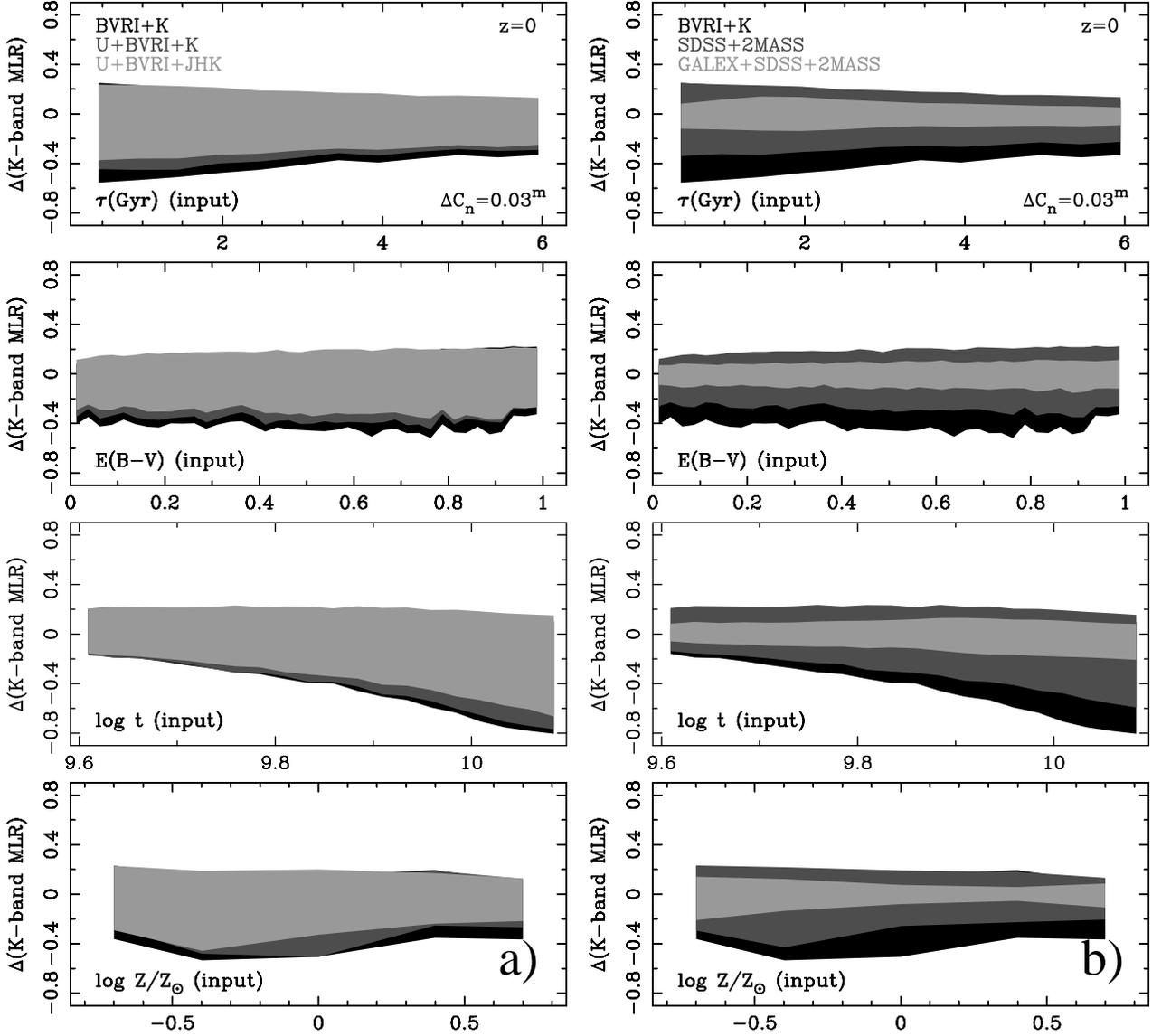}}
\end{figure}

\begin{figure}
\caption{{\bf a)} Uncertainties in the derived properties of
intermediate-redshift galaxies for the BVRI, U+BVRI, and UBVRI+K sets
assuming observing errors of 0.03\,mag. {\bf b)} $K$-band
mass-to-light ratio uncertainties for the BVRI+K, U+BVRI+K, and
U+BVRI+JHK sets assuming observing errors of 0.10\,mag. \label{fig7}}
\resizebox{0.93\hsize}{!}{\includegraphics*[133,235][485,556]{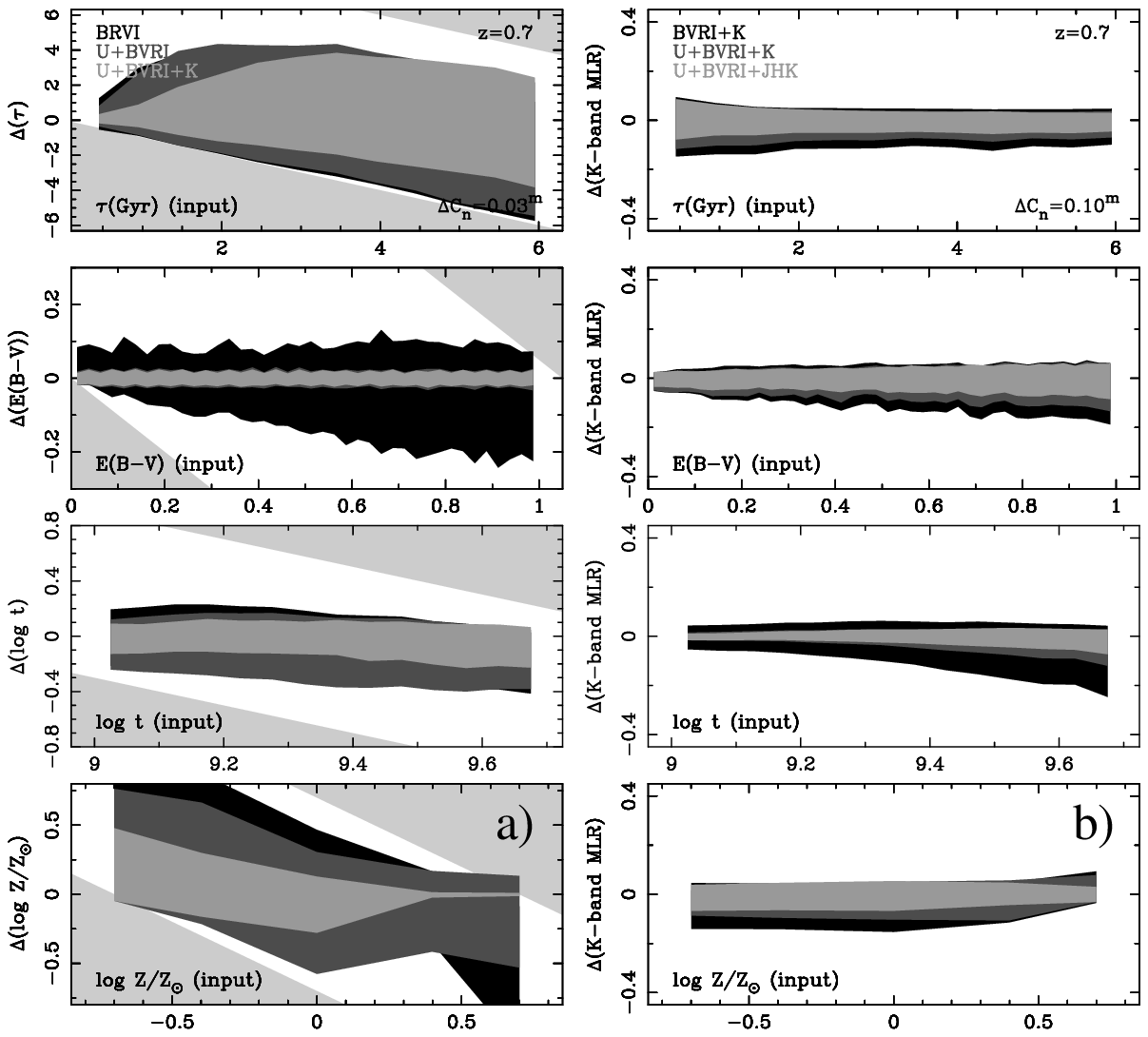}}
\end{figure}

\begin{figure}
\caption{{\bf a)} Uncertainties in the derived properties
of high-redshift galaxies for the BVRI, U+BVRI, and UBVRI+K
sets assuming observing errors of 0.03\,mag. {\bf b)} $K$-band
mass-to-light ratio uncertainties for the BVRI+K, U+BVRI+K, and
U+BVRI+JHK sets assuming observing errors of 0.10\,mag. \label{fig8}}
\resizebox{0.93\hsize}{!}{\includegraphics*[133,235][485,556]{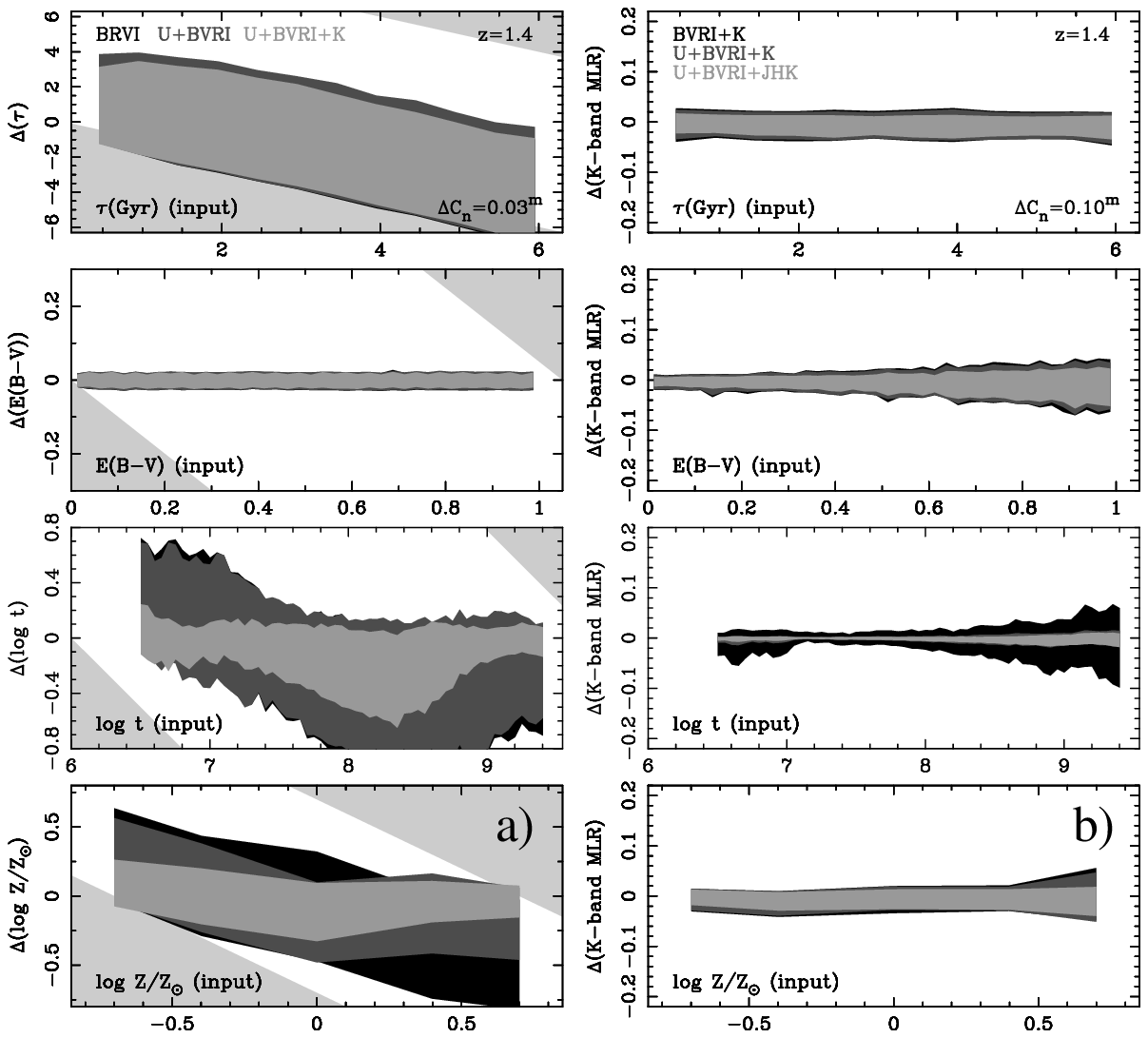}}
\end{figure}

With regard to the age determination, the uncertainties derived are
very similar for the BVRI and U+BVRI sets. However, for those
band-pass combinations including NIR data the ages derived are
strongly biased toward younger ages. The reason for this bias, which
also leads to significantly smaller age uncertainties, is the offset
in the ($J-H$) and ($H-K$) colors between the GISSEL99 and the
P\'{E}GASE model predictions (see Figure~\ref{fig1}) due to the
differences in the modelling of the upper RGB and AGB phases. In
particular, Figure~\ref{fig1}b shows that the P\'{E}GASE models are
$\sim$0.07\,mag redder in ($J-H$) and $\sim$0.04\,mag redder in
($H-K$) than the GISSEL99 models within the age range
4-12\,Gyr. Therefore, in order to compensate for this difference in
color, the best-fitting solution usually leads to younger ages, which
within this age range imply bluer colors. Because the differences in
the colors between the two models only occur in the NIR, the optical
colors predicted by the P\'{E}GASE models at these younger ages should
be bluer than those of the sample. Therefore, in order to compensate
for this effect, the age bias described above has to be accompanied by
strong biases in dust extinction and/or metallicity that would lead to
redder optical colors. Within the error intervals derived the total
degeneracy is dominated by the age-timescale and age-extinction
degeneracies.

\subsubsection{Dust Extinction} 
\label{sec4.4.3}

As we commented above (see also Figure~\ref{fig9}a) there is a strong
bias in dust extinction estimates toward higher extinction values when
NIR data are used. This bias, along with the metallicity bias
described in Section~\ref{sec4.4.4}, results in a global reddening of
the optical colors but a small change in the NIR colors of the
galaxies in the sample. On the other hand, at very high extinction
values the uncertainties are also biased by the upper limit in
$E(B-V)$ imposed during the data-models comparison procedure (see
Table~\ref{table1}). The mean uncertainties derived, both in age, dust
extinction, and metallicity are very similar to those obtained by
using the GISSEL99 models.

\subsubsection{Metallicity}
\label{sec4.4.4}

The distribution of the uncertainty in metallicity shown in
Figure~\ref{fig9}a indicates that a strong bias toward higher
metallicity values is present when NIR data are available. As we
commented in Section~\ref{sec4.4.2}, this bias is probably related
with the age bias and the differences in the NIR colors predicted by
the two sets of models. As in the case of GISELL~99 models, the
comparison with the P\'{E}GASE models leads to a clear dependence of
the uncertainty with the metallicity value itself, with smaller
uncertainties at very high metallicities (see Section~\ref{sec4.1.4}).

\subsubsection{Stellar Mass} 
\label{sec4.4.5}

\begin{figure}
\caption{Mean timescale, age, dust extinction, metallicity (panel {\bf
a}), and $K$-band mass-to-light ratio (panel {\bf b}) uncertainties
derived using the P\'{E}GASE evolutionary synthesis models. In panel {\bf
a} we show the results for the BVRI, U+BVRI, and UBVRI+K sets, and in
panel {\bf b} those for the BVRI+K, U+BVRI+K, and U+BVRI+JHK sets. In
both cases observing errors of 0.03\,mag have been adopted. \label{fig9}}
\resizebox{0.93\hsize}{!}{\includegraphics*[133,235][485,556]{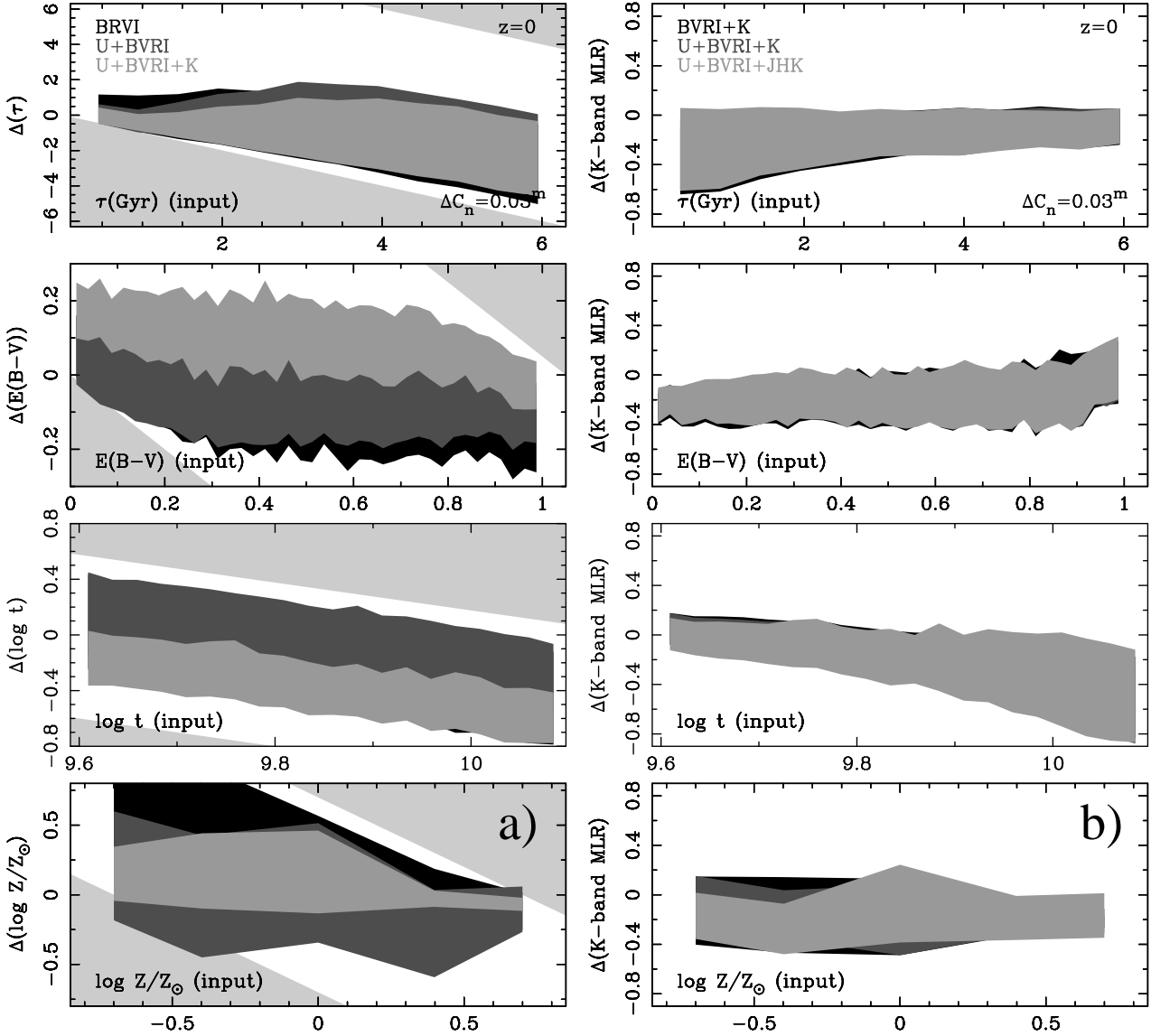}}
\end{figure}

The results shown in Figure~\ref{fig9}b with regard to the $K$-band
mass-to-light ratios mainly reflect the biases in the galaxy property
determination, with the stellar masses derived systematically smaller
than the input values. This is due (1) to the bias toward younger ages
described in Section~\ref{sec4.4.2} and (2) to the higher $K$-band
luminosity per unit mass of the P\'{E}GASE models compared with the
GISSEL99 models (see Figure~\ref{fig1}b). Because of the stronger bias
in age, the mean uncertaintines in the $K$-band mass-to-light ratio
are smaller than those obtained using the GISSEL99 models (see
Section~\ref{sec4.1.5}). Finally, Figure~\ref{fig9}b shows that the
mass-to-light ratio uncertainty becomes higher at older ages and lower
timescale values. This behavior, which is also present in the case of
the GISSEL99 models (see Section~\ref{sec4.1.5}), is due to the
progressive increase in the rate of change of the $K$-band
mass-to-light ratio with log~$t$ (see Figure~\ref{fig1}) accompanied
by a small dependence of the age uncertainty (in log~$t$ scale) with
the value of the age itself.

\section{Conclusions}
\label{sec5}

In this study we have analyzed the dependence of the uncertainties and
degeneracies in the galaxy properties upon different parameters: (1)
the combination of bands available, (2) the observing errors, and (3)
the galaxy properties themselves (including redshift).

Here we summarize our main results and point out some directions for
the optimization of galaxy evolution studies using broad-band
photometry data. We describe separately the nearby, intermediate, and
high-redshift cases.

{\bf Nearby galaxies:} In order to determine the star formation
history, age, and dust extinction of nearby galaxies with relatively
small uncertainties the use of $U$-band data is fundamental. The
availability of $K$-band data also allows a reduction in the
uncertainty in the age and metallicity of the stellar population, but
the use of additional $J$ and $H$-band data is largely redundant. The
use of the $K$-band data is unfortunately limited by the existence of
large uncertainties in the modeling of the $K$-band luminosities and
NIR colors of stellar populations. The most significant reduction in
the age and $K$-band mass-to-light ratio uncertainty is achieved when
UV data are used. The poor treatment of the post-AGB and ``extreme''
HB phases by the existing evolutionary synthesis models introduce,
however, an additional uncertainty during the data-model comparison,
which is particularly important in the case of very old stellar
populations. For the same number of observing bands, the availability
of wider wavelength baselines results in lower uncertainties. Both the
formation timescale and $K$-band mass-to-light ratio uncertainties are
larger when the corresponding values for these properties are
larger. On the other hand, the metallicity uncertainty decreases with
the value of the metallicity itself due to the distinct photometric
evolution of high-metallicity core Helium burning stars.

A complete description of the physical reasons behind these
conclusions and of the degeneracies responsible for the uncertainties
described above are given in Section~\ref{sec4.1} (see also
Section~\ref{sec4.4}).

{\bf Intermediate-redshift galaxies:} The star formation history of
intermediate-redshift galaxies can be derived with worse precision
than in nearby galaxies because their stellar populations are
younger. The age uncertainty is smaller than in the nearby-galaxies
case and shows a strong bias toward younger ages. A significant
reduction of this bias and of the mean uncertainties is achieved when
NIR data are used, especially if all three $J$, $H$, and $K$-band data
are available. The dust-extinction uncertainty is larger for larger
values of the dust extinction itself. The use of $U$-band data
provides an important reduction of this dependence and of the mean
dust-extinction uncertainty. If $U$-band data are available the use of
additional UV data do not provide much more information about the
galaxy properties. The use of NIR data ($J$, $H$, and $K$-band data)
significantly reduces the uncertainty in the metallicity of the
galaxy. The absolute and relative uncertainties in the galaxy $K$-band
mass-to-light ratio are smaller than those derived for nearby
galaxies. However, the fact that the $K$ filter now covers the
rest-frame $J$-band leads to a larger uncertainty associated with the
IMF and with the parameterization of the galaxy star formation history
and, consequently, to a larger stellar mass uncertainty. For a more
detailed description see Section~\ref{sec4.2}.

{\bf High-redshift galaxies:} As expected, the bias and mean
uncertainty in the determination of the timescale for the galaxy
formation are even larger in this case that in the nearby or
intermediate-redshift galaxies. The age of the stellar population is
derived with a large uncertainty, only reduced when $JHK$ data are
available. The dust-extinction in these galaxies can be derived to a
very high accuracy even when only optical data are available. The use
of $JHK$ data is fundamental in order to improve both the age and
metallicity determinations. A complete description of the
uncertainties and degeneracies between these properties is given in
Section~\ref{sec4.3}.

Some of the conclusions drawn above can also be found through the
literature expressed in a qualitative way. However, this work
constitutes the first systematic and quantitative study on the
optimization of broad-band photometry for studies on the evolution of
galaxies. It is important to note that the application of these
results to future galaxy surveys can help to reduce the uncertainty in
the derivation of the galaxy physical properties, sometimes by
weakening a particular degeneracy but most of the time by decreasing
the intervals over which this degeneracy takes place. In
Table~\ref{table3} we have summarized the mean uncertainties in the
galaxy stellar population properties derived in this paper considering
different redshifts, sets of observables, and observing errors.

Our results are directly applicable to spectrophotometric surveys like
the SDSS and surveys looking for emission-line galaxies at fixed
redshifts (Martin, Lotz \& Ferguson 2000; Moorwood, van der Werf, Cuby
\& Oliva 2000; Iwamuro et al$.$ 2000; Pascual et al$.$ 2001; Zamorano
et al$.$, in preparation). However, in the case of the blind-redshift
surveys a comparison between our results and those from previous
studies on the optimization of the photometric-redshifts technique
(Kodama et al$.$ 1999; Bolzonella et al$.$ 2000; Mobasher \& Mazzei
2000; Wolf et al$.$ 2001) is still needed.

\section{Future applications}
\label{sec6}

The results summarized above demonstrate that the design of galaxy
evolution studies based only on qualitative, intuitive ideas may lead
(in some cases avoidably) to large uncertainties.

Because of this we intend to apply this work to the design of future
projects on galaxy evolution estimating optimal sets of observables
and required signal-to-noise ratios. Although in this work we have
only considered broad-band filters, this procedure is easily
generalizable to combinations of broad, medium, and narrow-band
filters from the far-UV to the near-infrared. In addition, the
combined use of the procedure here described with state-of-the-art
radiative transfer and dust models (Popescu et al$.$ 2000) will allow
us to extend this range up to sub-millimeter wavelengths.

Beyond the results shown in this paper, we can also derive, upon
request, the uncertainties and degeneracies in the galaxy properties
for a given combination of filters, observing errors, and galaxy
redshift.

\acknowledgements We would like to thank the referee for the useful
comments and suggestions that have significantly improved the quality
of this paper. We are also very grateful to A$.$ G$.$ Bruzual for
providing the GISSEL99 evolutionary synthesis models, P$.$G$.$ P\'erez
Gonz\'alez, N$.$ Cardiel and O$.$ Pevunova for stimulating
conversations, and C$.$ S\'anchez Contreras for carefully reading the
manuscript and making some useful remarks. A$.$ G$.$ de P$.$
acknowledges the financial support from NASA through a Long Term Space
Astrophysics grant to B$.$ F$.$ M.


\begin{deluxetable}{lcc}
\tablecaption{Range in galaxy properties from which synthetic galaxy
colors of the sample were generated and range in the model parameters for the data-models comparison.\label{table1}}
\tablewidth{0pt}
\tablehead{\colhead{}      &  \colhead{Sample} & \colhead{Models}}
\startdata
$E(B-V)_{\mathrm{stellar}}$         &   0.00-1.00\,mag & 0.00-1.05\,mag \\
$\tau$           &    0.2-6\,Gyr    & 0.01-10\,Gyr   \\
Age ($z$=0)      &    4-12\,Gyr     & 1-15\,Gyr      \\
\ \ \  ($z$=0.7) &    1-5\,Gyr      & 0.5-8\,Gyr     \\
\ \ \  ($z$=1.4) &    3\,Myr-3\,Gyr & 1\,Myr-6\,Gyr  \\
Metallicity (Z)  &  0.2, 0.4, 1.0, 2.5, 5$\times$Z$_{\odot}$ & 0.2, 0.4, 1.0, 2.5, 5$\times$Z$_{\odot}$ \\
\enddata
\end{deluxetable}

\begin{deluxetable}{lcc}
\tablecaption{Definition of the different combinations of filters analyzed \label{table2}}
\tablewidth{0pt}
\tablehead{\colhead{Bands} & \colhead{\# of bands} & \colhead{Alias}}
\startdata
$B,V,R_{C},I_{C}$                & 4  & BVRI\\
$U,B,V,R_{C},I_{C}$              & 5  & U+BVRI\\
$U,B,V,R_{C},I_{C},K$            & 6  & U+BVRI+K\\
$U,B,V,R_{C},I_{C},J,H,K$        & 8  & U+BVRI+JHK\\ 
$B,V,R_{C},I_{C},J,H,K$          & 7  & BVRI+JHK\\
$B,V,R_{C},I_{C},K$              & 5  & BVRI+K\\
$U,V,I_{C},J,K$                  & 5  & UVIJK\\
$u',g',r',i',z'$                 & 5  & SDSS\\
$u',g',r',i',z',J,H,K$           & 8  & SDSS+2MASS\\
FUV,NUV,$u',g',r',i',z',J,H,K$   & 10 & GALEX+SDSS+2MASS\\
\enddata
\end{deluxetable}


\begin{table}
\caption{Mean 1$\sigma$ uncertainties in the derived properties \label{table3}}
\begin{small} 
\begin{tabular}{lccccccccccc}
\tableline
\tableline
  {\bf Set}          & Property & Unit  & \multicolumn{3}{c}{$z$=0} & \multicolumn{3}{c}{$z$=0.7} & \multicolumn{3}{c}{$z$=1.4} \\
\tableline
 & \multicolumn{2}{r}{$\Delta C_n$ (mag) =}&  0.03 & 0.07 & 0.10    &      0.03 & 0.07 & 0.10     &      0.03 & 0.07 & 0.10     \\
\tableline
      BVRI           & $E(B-V)$ &  mag  &     0.16 & 0.17 & 0.17    &      0.11 & 0.18 & 0.20     &      0.02 & 0.04 & 0.06     \\
                     &  $\tau$  &  Gyr  &     2.50 & 2.87 & 3.07    &      2.89 & 3.10 & 3.32     &      2.91 & 3.33 & 3.42     \\   
                     & log~$t$  &  dex  &     0.36 & 0.37 & 0.38    &      0.23 & 0.31 & 0.35     &      0.45 & 0.71 & 0.84     \\
                     & log~$Z$  &  dex  &     0.33 & 0.33 & 0.33    &      0.44 & 0.46 & 0.45     &      0.37 & 0.45 & 0.48     \\
&(M/L)$_{K}$& M$_{\odot}$/L$_{K,\odot}$ &     0.32 & 0.41 & 0.47    &      0.06 & 0.08 & 0.09     &      0.03 & 0.05 & 0.06     \\
\multicolumn{12}{c}{}\\
     U+BVRI          & $E(B-V)$ &  mag  &     0.14 & 0.18 & 0.19    &      0.02 & 0.04 & 0.07     &      0.02 & 0.04 & 0.06     \\
                     &  $\tau$  &  Gyr  &     2.27 & 2.69 & 3.00    &      3.10 & 3.10 & 3.11     &      3.11 & 3.34 & 3.33     \\   
                     & log~$t$  &  dex  &     0.34 & 0.37 & 0.38    &      0.23 & 0.30 & 0.34     &      0.45 & 0.68 & 0.81     \\
                     & log~$Z$  &  dex  &     0.32 & 0.41 & 0.45    &      0.38 & 0.51 & 0.54     &      0.29 & 0.37 & 0.38     \\
&(M/L)$_{K}$& M$_{\odot}$/L$_{K,\odot}$ &     0.32 & 0.41 & 0.47    &      0.05 & 0.08 & 0.09     &      0.03 & 0.05 & 0.06     \\
\multicolumn{12}{c}{}\\
     U+BVRI+K        & $E(B-V)$ &  mag  &     0.09 & 0.14 & 0.17    &      0.02 & 0.04 & 0.05     &      0.02 & 0.04 & 0.05     \\
                     &  $\tau$  &  Gyr  &     2.19 & 2.75 & 3.03    &      2.29 & 2.95 & 3.19     &      2.79 & 3.00 & 3.08     \\   
                     & log~$t$  &  dex  &     0.28 & 0.35 & 0.38    &      0.13 & 0.21 & 0.24     &      0.20 & 0.28 & 0.33     \\
                     & log~$Z$  &  dex  &     0.11 & 0.20 & 0.26    &      0.15 & 0.22 & 0.26     &      0.17 & 0.29 & 0.34     \\
&(M/L)$_{K}$& M$_{\odot}$/L$_{K,\odot}$ &     0.26 & 0.33 & 0.35    &      0.03 & 0.06 & 0.06     &      0.01 & 0.02 & 0.03     \\
\multicolumn{12}{c}{}\\
     U+BVRI+JHK      & $E(B-V)$ &  mag  &     0.08 & 0.14 & 0.17    &      0.02 & 0.03 & 0.05     &      0.02 & 0.03 & 0.04     \\
                     &  $\tau$  &  Gyr  &     2.12 & 2.70 & 2.96    &      2.35 & 3.07 & 3.35     &      3.19 & 3.44 & 3.44     \\   
                     & log~$t$  &  dex  &     0.26 & 0.36 & 0.39    &      0.09 & 0.17 & 0.22     &      0.12 & 0.23 & 0.29     \\
                     & log~$Z$  &  dex  &     0.09 & 0.20 & 0.26    &      0.10 & 0.17 & 0.22     &      0.06 & 0.17 & 0.23     \\
&(M/L)$_{K}$& M$_{\odot}$/L$_{K,\odot}$ &     0.24 & 0.35 & 0.37    &      0.02 & 0.04 & 0.05     &      0.01 & 0.01 & 0.02     \\
\multicolumn{12}{c}{}\\
      BVRI+JHK       & $E(B-V)$ &  mag  &     0.10 & 0.17 & 0.18    &      0.04 & 0.10 & 0.13     &      0.02 & 0.03 & 0.05     \\
                     &  $\tau$  &  Gyr  &     2.53 & 2.97 & 3.14    &      2.38 & 3.08 & 3.30     &      3.22 & 3.47 & 3.45     \\   
                     & log~$t$  &  dex  &     0.32 & 0.38 & 0.40    &      0.12 & 0.22 & 0.28     &      0.13 & 0.24 & 0.29     \\
                     & log~$Z$  &  dex  &     0.13 & 0.23 & 0.28    &      0.11 & 0.20 & 0.25     &      0.07 & 0.19 & 0.25     \\
&(M/L)$_{K}$& M$_{\odot}$/L$_{K,\odot}$ &     0.29 & 0.37 & 0.39    &      0.03 & 0.05 & 0.06     &      0.01 & 0.02 & 0.02     \\
\multicolumn{12}{c}{}\\
      BVRI+K         & $E(B-V)$ &  mag  &     0.12 & 0.17 & 0.18    &      0.08 & 0.14 & 0.16     &      0.02 & 0.04 & 0.05     \\
                     &  $\tau$  &  Gyr  &     2.67 & 3.15 & 3.35    &      2.32 & 2.88 & 3.06     &      2.72 & 2.92 & 3.03     \\   
                     & log~$t$  &  dex  &     0.34 & 0.37 & 0.38    &      0.16 & 0.25 & 0.30     &      0.22 & 0.30 & 0.34     \\
                     & log~$Z$  &  dex  &     0.15 & 0.23 & 0.28    &      0.20 & 0.29 & 0.31     &      0.26 & 0.39 & 0.44     \\
&(M/L)$_{K}$& M$_{\odot}$/L$_{K,\odot}$ &     0.30 & 0.34 & 0.36    &      0.05 & 0.08 & 0.09     &      0.02 & 0.03 & 0.03     \\
\multicolumn{12}{c}{}\\
      UVIJK          & $E(B-V)$ &  mag  &     0.10 & 0.16 & 0.19    &      0.02 & 0.04 & 0.07     &      0.02 & 0.03 & 0.05     \\
                     &  $\tau$  &  Gyr  &     2.33 & 2.87 & 3.14    &      2.54 & 3.21 & 3.49     &      3.14 & 3.41 & 3.42     \\   
                     & log~$t$  &  dex  &     0.30 & 0.36 & 0.38    &      0.11 & 0.18 & 0.23     &      0.16 & 0.27 & 0.33     \\
                     & log~$Z$  &  dex  &     0.13 & 0.24 & 0.29    &      0.11 & 0.18 & 0.23     &      0.08 & 0.19 & 0.24     \\
&(M/L)$_{K}$& M$_{\odot}$/L$_{K,\odot}$ &     0.29 & 0.36 & 0.38    &      0.02 & 0.04 & 0.05     &      0.01 & 0.02 & 0.02     \\
\multicolumn{12}{c}{}\\
       SDSS          & $E(B-V)$ &  mag  &     0.10 & 0.14 & 0.17    &      0.02 & 0.06 & 0.09     &      0.03 & 0.05 & 0.07     \\
                     &  $\tau$  &  Gyr  &     2.28 & 2.73 & 2.93    &      2.93 & 3.11 & 3.23     &      2.94 & 3.05 & 2.97     \\   
                     & log~$t$  &  dex  &     0.31 & 0.39 & 0.41    &      0.18 & 0.26 & 0.31     &      0.39 & 0.61 & 0.74     \\
                     & log~$Z$  &  dex  &     0.21 & 0.31 & 0.35    &      0.40 & 0.44 & 0.45     &      0.25 & 0.32 & 0.37     \\
&(M/L)$_{K}$& M$_{\odot}$/L$_{K,\odot}$ &     0.30 & 0.38 & 0.40    &      0.04 & 0.06 & 0.08     &      0.02 & 0.04 & 0.05     \\
\multicolumn{12}{c}{}\\
      SDSS+2MASS     & $E(B-V)$ &  mag  &     0.07 & 0.14 & 0.16    &      0.02 & 0.04 & 0.05     &      0.02 & 0.03 & 0.04     \\
                     &  $\tau$  &  Gyr  &     2.06 & 2.65 & 2.89    &      2.36 & 3.07 & 3.34     &      3.20 & 3.41 & 3.39     \\   
                     & log~$t$  &  dex  &     0.25 & 0.36 & 0.39    &      0.09 & 0.16 & 0.21     &      0.12 & 0.21 & 0.26     \\
                     & log~$Z$  &  dex  &     0.08 & 0.19 & 0.25    &      0.10 & 0.16 & 0.21     &      0.08 & 0.17 & 0.23     \\
&(M/L)$_{K}$& M$_{\odot}$/L$_{K,\odot}$ &     0.24 & 0.35 & 0.37    &      0.02 & 0.04 & 0.05     &      0.01 & 0.01 & 0.02     \\
\multicolumn{12}{c}{}\\
  GALEX+SDSS+        & $E(B-V)$ &  mag  &     0.04 & 0.09 & 0.12    &      0.01 & 0.02 & 0.02     &      0.01 & 0.02 & 0.03     \\
 +2MASS              &  $\tau$  &  Gyr  &     1.51 & 2.26 & 2.60    &      2.01 & 2.93 & 3.26     &      2.92 & 3.63 & 3.67     \\
                     & log~$t$  &  dex  &     0.10 & 0.19 & 0.24    &      0.08 & 0.13 & 0.17     &      0.09 & 0.16 & 0.21     \\
                     & log~$Z$  &  dex  &     0.05 & 0.16 & 0.22    &      0.11 & 0.16 & 0.20     &      0.03 & 0.08 & 0.14     \\
&(M/L)$_{K}$& M$_{\odot}$/L$_{K,\odot}$ &     0.10 & 0.23 & 0.28    &      0.02 & 0.03 & 0.04     &      0.01 & 0.01 & 0.01     \\
\tableline
\end{tabular}
\vspace{-0.7cm} 
\end{small} 
\end{table}

\end{document}